\documentclass[11pt]{article}
\pdfoutput=1 
\usepackage{shorthand}
\usepackage{mathtools}
\usepackage{booktabs}
\usepackage[english]{babel}
\usepackage{amsmath,amssymb,amsbsy,amstext, amsthm, simplewick, amsfonts,braket}
\usepackage{graphicx}
\usepackage[small]{caption}
\usepackage{siunitx}
\usepackage{upgreek}
\usepackage{framed}
\usepackage{wrapfig}
\usepackage{multirow}
\usepackage{bbm}
\usepackage[numbers,sort&compress]{natbib}
\usepackage[svgnames,dvipsnames,x11names]{xcolor}
\usepackage[utf8x]{inputenc}
\usepackage{selinput}
\usepackage{bm}
\usepackage{float}
\usepackage{dsfont}
\usepackage{caption}
\usepackage{subcaption}
\usepackage{sidecap}
\usepackage{longtable}
\usepackage{anyfontsize}

\usepackage{epstopdf}
\usepackage{cancel}
\usepackage{tcolorbox}
\usepackage{latexsym,amsmath,amssymb,epsfig}
\usepackage{braket}
\usepackage{tensor}
\usepackage{tocloft}

\usepackage{tikz-cd}

\usepackage{ytableau}
\ytableausetup{centertableaux,boxsize=.8em}
\usepackage{genyoungtabtikz}
\Yboxdim{14pt} 
\Ylinethick{.9pt}
\newcommand*{\medotimes}{\raisebox{-0.05ex}{\scalebox{1.25}{$\otimes$}}}

\usepackage{tcolorbox}
\definecolor{greyish2}{rgb}{.96,.96,.96}

\def\xyma{\xymatrix@M.7em}
\def\xymas{\xymatrix@M.1em}

\newcommand{\Comment}[1]{{}}
\definecolor{darkblue}{rgb}{0.15,0.35,0.55}
\definecolor{reddish}{rgb}{0.65, 0.2, 0.2}
\definecolor{darkgreen}{RGB}{50,150,0}
\definecolor{greyish2}{rgb}{.96,.96,.96}
\usepackage[linktocpage=true]{hyperref}
\hypersetup{
colorlinks=true,
citecolor=darkblue,
linkcolor=reddish,
urlcolor=darkblue,
pdfauthor={},
pdftitle={},
pdfsubject={}
}

\DeclareFontFamily{OT1}{rsfs10}{}
\DeclareFontShape{OT1}{rsfs10}{m}{n}{ <-> rsfs10 }{}
\DeclareMathAlphabet{\mathscript}{OT1}{rsfs10}{m}{n}

\def\gsim{ \lower .75ex \hbox{$\sim$} \llap{\raise .27ex \hbox{$>$}} }
\def\lsim{ \lower .75ex \hbox{$\sim$} \llap{\raise .27ex \hbox{$<$}} }
\def\be{\begin{equation}}
\def\ee{\end{equation}}
\def\bea{\begin{eqnarray}}
\def\eea{\end{eqnarray}}
\def\KL{K\"all\'en--Lehmann }

\newcommand{\baaa}{\begin{eqnarray}}
\newcommand{\eaaa}{\end{eqnarray}}

\newcommand{\rd}{{\rm d}}

\newcommand{\tr}{\text{tr}}

\usepackage[letterpaper,margin=1in]{geometry}

\usepackage{tikz}
\usetikzlibrary{decorations}
\pgfdeclaredecoration{complete sines}{initial}
{
    \state{initial}[
        width=+0pt,
        next state=upsine,
        persistent precomputation={\pgfmathsetmacro\matchinglength{
            \pgfdecoratedinputsegmentlength / int(\pgfdecoratedinputsegmentlength/\pgfdecorationsegmentlength)}
            \setlength{\pgfdecorationsegmentlength}{\matchinglength pt}
        }] {}
    \state{upsine}[width=\pgfdecorationsegmentlength,next state=downsine]{
        \pgfpathsine{\pgfpoint{0.25\pgfdecorationsegmentlength}{0.5\pgfdecorationsegmentamplitude}}
        \pgfpathcosine{\pgfpoint{0.25\pgfdecorationsegmentlength}{-0.5\pgfdecorationsegmentamplitude}}
    }
    \state{downsine}[width=\pgfdecorationsegmentlength,next state=upsine]{
        \pgfpathsine{\pgfpoint{0.25\pgfdecorationsegmentlength}{-0.5\pgfdecorationsegmentamplitude}}
        \pgfpathcosine{\pgfpoint{0.25\pgfdecorationsegmentlength}{0.5\pgfdecorationsegmentamplitude}}
}
    \state{final}{}
}

\definecolor{greyish}{rgb}{.90,.90,.90}
\definecolor{greyish2}{rgb}{.96,.96,.96}
\usepackage{xcolor,colortbl}
\usepackage{tcolorbox}

\usepackage[all]{xy}


\numberwithin{equation}{section}

\begin{document}

%
\renewcommand{\thefootnote}{\fnsymbol{footnote}}
\vspace{0truecm}
\thispagestyle{empty}

\begin{center}
{\fontsize{20}{24} \bf Impossible Symmetries\\[5pt] and
Conformal Gravity}
\end{center}

\vspace{.025truecm}

\begin{center}
{\fontsize{12.7}{18}\selectfont
Kurt Hinterbichler,${}^{\rm a}$\footnote{\texttt{\href{mailto:kurt.hinterbichler@case.edu}{kurt.hinterbichler@case.edu}}}
Austin Joyce,${}^{\rm b}$\footnote{\texttt{\href{mailto:austinjoyce@uchicago.edu}{austinjoyce@uchicago.edu}}}
and Gr\'egoire Mathys,${}^{\rm c,\, d}$\footnote{\texttt{\href{mailto:gregoire.mathys@epfl.ch}{gregoire.mathys@epfl.ch}}} 
}
\end{center}

\vspace{.2truecm}

 \centerline{{\it ${}^{\rm a}$CERCA, Department of Physics,}}
 \centerline{{\it Case Western Reserve University, 10900 Euclid Ave, Cleveland, OH 44106, USA}} 
 
\vspace{.3cm}

   \centerline{{\it ${}^{\rm b}$Kavli Institute for Cosmological Physics, Department of Astronomy and Astrophysics}}
 \centerline{{\it University of Chicago, Chicago, IL 60637, USA} } 
 
\vspace{.3cm}

\centerline{{\it ${}^{\rm c}$Department of Physics,}}
 \centerline{{\it Cornell University, Ithaca, NY 14850, USA} } 

\vspace{.3cm}

\centerline{{\it ${}^{\rm d}$Fields and Strings Laboratory, Institute of Physics}}
 \centerline{{\it Ecole Polytechnique Fédérale de Lausanne (EPFL)} } 
 \centerline{{\it Route de la Sorge, CH-1015 Lausanne, Switzerland}}
 \vspace{.25cm}

\vspace{.3cm}
\begin{abstract}
\noindent
We explore the physics of relativistic gapless phases defined by a mixed anomaly between two generalized conserved currents. The gapless modes can be understood as Goldstone modes arising from the nonlinear realization of (generically higher-form) symmetries arising from these currents. 
In some cases, the anomaly cannot be reproduced by any local and unitary theory, indicating that the corresponding symmetries are impossible, in the sense that they cannot appear in a Lorentzian physical system.
We give a general construction and illustrate it with several examples. Most notably, we study conformal gravity from this perspective, describing the higher-form symmetries of the linear theory and showing how it can be understood in terms of anomalies. Along the way we clarify some aspects of electric-magnetic duality in linear conformal gravity.

\end{abstract}

\newpage

\setcounter{page}{2}
\setcounter{tocdepth}{2}
\tableofcontents
\newpage
\renewcommand*{\thefootnote}{\arabic{footnote}}
\setcounter{footnote}{0}




\section{Introduction}
\label{sec:intro}

Symmetries help to delineate the boundaries between the possible and the impossible. For example, selection rules tell us that transitions between certain states cannot happen. As we will see, sometimes symmetries {\it themselves} are impossible: certain patterns of symmetries cannot be unitarily realized by any low-energy effective field theory (EFT).

One of the goals of physics is to understand the possible phases that the degrees of freedom that make up our universe can be in.
In this pursuit, the Landau paradigm and effective field theory  serve as powerful organizing principles which allow us to describe a wide variety of phenomena using a unified language. 
From this perspective, we classify phases both by their symmetries and by how these symmetries are realized on the light degrees of freedom of the system.
Recently, the horizons of the Landau paradigm have been significantly broadened by a commensurate expansion of our conceptions of symmetries themselves. It has been understood that ordinary symmetries admit vast generalizations in many directions, from higher forms~\cite{Kapustin:2013uxa,Gaiotto:2014kfa} to higher groups~\cite{Cordova:2018cvg,Benini:2018reh} and even to non-invertible and categorical settings~\cite{Verlinde:1988sn,Chang:2018iay,Komargodski:2020mxz,Choi:2021kmx,Kaidi:2021xfk,GarciaEtxebarria:2022jky,Shao:2023gho}. These formal structures have already proven to be useful in phenomenological applications~\cite{Choi:2022jqy,Cordova:2022ieu,Cordova:2022fhg,Cordova:2023her,Choi:2023pdp,Reece:2023iqn,Agrawal:2023sbp} and
the interplay between these structures and the principles of effective field theory is just beginning to be explored~\cite{Grozdanov:2016tdf,Lake:2018dqm,Hofman:2018lfz,Armas:2018zbe,Armas:2018atq,Glorioso:2018kcp,Delacretaz:2019brr,Thorngren:2023ple,Armas:2023tyx,Armas:2019sbe,Armas:2022wvb,McGreevy:2022oyu,Vardhan:2022wxz,Landry:2022nog,Das:2022fho,Hinterbichler:2022agn,Benedetti:2023ipt}. In this regard, understanding what symmetries can consistently appear in effective descriptions is essential.,

In light of these developments,
it is worthwhile to reconsider what actually defines an effective field theory, or a phase of matter, in order to fully realize the potential of the Landau paradigm.  A natural operational definition is to specify the local conserved currents, along with their compatibility, as expressed by the structure of 't Hooft anomalies in the presence of background sources. This viewpoint is surprisingly powerful, making it possible to describe phenomena as varied as superfluids and superconductors~\cite{Delacretaz:2019brr,Thorngren:2023ple,Armas:2023tyx} to gravity~\cite{Benedetti:2021lxj,Hinterbichler:2022agn,Benedetti:2023ipt,Hull:2023iny} in a unified language. 
Given these successes, one might imagine that any pattern of symmetries and anomalies can be realized by {\it some} quantum field theory. But, quite remarkably, it has long been known that certain apparently consistent effective field theories do not admit ultraviolet completions, satisfying certain requirements~\cite{Adams:2006sv}.\footnote{A thematically related line of inquiry is the ``swampland" program~\cite{Palti:2019pca}, which is concerned with the constraints placed on EFTs by the requirement that they can consistently be ultraviolet completed by quantum gravity. A difference, however, is that swampland constraints are not supposed to be visible within the EFT itself.}

In this paper, we explore this ultraviolet-infrared (UV-IR) interplay in the context of the {\it symmetries} that arise in a low-energy field theory. Following~\cite{Delacretaz:2019brr,Hinterbichler:2022agn}, we define an EFT via its conserved currents.  We allow for these currents to be generalized (i.e., they can be operators in arbitrary mixed-symmetry representations of the Lorentz group, satisfying general conservation conditions involving one or more derivatives acting in various ways).  We consider situations where a relativistic system possesses two of these generalized currents in the IR: one current, $K$, which we call ``magnetic,'' and one current, $J$, which we call ``electric''.  These currents can have a mixed  anomaly between them---one of the two conservation conditions must necessarily be broken in the presence of background gauge fields for the currents, or equivalently, conservation must be broken at coincident points in correlators.  By assuming such a non-vanishing mixed anomaly between the currents,
we show that the system must necessarily be in a gapless phase. The properties of the massless degrees of freedom are then determined by the structure of the currents and their conservation conditions.   We will see that sometimes these gapless modes are necessarily non-unitary, which means that the corresponding 
combinations of conserved currents and anomalies cannot be realized by any unitary field theory. This then implies that these symmetries are impossible in the sense that they cannot arise in the low-energy description of any unitary and Lorentz invariant microphysical theory.
These symmetries need not be exact symmetries of the underlying microphysical theory---and indeed in general we would not expect them to be. Rather, they can be emergent at low energies. Remarkably, even this possibility is not allowed.

In order to illustrate the features of the construction, we consider several examples where the gapless mode is a scalar degree of freedom, and show how they can be understood in a uniform way. In addition to the familiar example of an ordinary superfluid~\cite{Delacretaz:2019brr}, and the slightly more exotic galileon superfluid considered in~\cite{Hinterbichler:2022agn}, we also consider the theory of a higher-derivative conformal scalar. This theory is structurally similar to conformal gravity, and the corresponding gapless phase is necessarily non-unitary, providing a simple example of a theory with ``impossible" symmetries.

Perhaps one of the more interesting applications---and our original motivating example---is to conformal gravity. Weyl-squared gravity is a four-derivative theory of gravity which propagates more degrees of freedom than Einstein gravity. 
In four spacetime dimensions it is Weyl invariant and is known as conformal gravity~\cite{Weyl:1918ib,Weyl:1919fi,Weyl:1918pdp,Bach:1921zdq}.
Famously, some of its propagating modes necessarily have a wrong sign kinetic term (as can be seen explicitly around an (A)dS background~\cite{Maldacena:2011mk,Deser:2012qg,Deser:2012euu,Joung:2014aba}) so that the theory is non-unitary. We will describe the higher-form symmetries of linearized Weyl-squared gravity (which is Weyl and conformally invariant in all dimensions, in contrast to its nonlinear version), and show how it can be thought of as a relativistic gapless phase with a particular structure of symmetries and anomalies.
Unsurprisingly, this structure cannot be realized by any unitary theory, showing more abstractly that (linearized) conformal gravity cannot arise as the low-energy description of a unitary QFT. 
We also consider conserved charges in this theory, and elucidate the structure of its electric-magnetic duality in $D=4$.

These results form a version of a UV-IR connection. Given our ignorance of UV physics as low-energy observers, we might imagine that it is possible for the microscopic theory that describes our world to have essentially any symmetries.\footnote{Presumably these symmetries would be only approximate at some energy scale, given the expectation that quantum gravity breaks all global symmetries.} The examples we study give obstructions to certain symmetries being present in the UV, assuming unitarity and relativistic invariance of the IR theory.  
We expect that these insights will help to more broadly understand the systematics of generalized symmetries in the context of EFT.

\vskip4pt
\noindent
{\bf Outline:} In Section~\ref{sec:general} we describe the general construction of relativistic gapless phases in terms of a pair of conserved currents with a mixed anomaly. 
In Section~\ref{sec:scalarsec} we illustrate this construction with several scalar field examples, including one that is conceptually similar to conformal gravity. In Section~\ref{sec:gravsec} we review the higher-form symmetries of linearized Einstein gravity then describe the higher-form symmetries of linear conformal gravity and cast it as a gapless phase. We conclude in Section~\ref{sec:conclusion}. In Appendix~\ref{app:confgrav} we briefly review some of the salient features of both nonlinear and linear conformal gravity. In Appendix~\ref{ap:Spec} we describe the spectral decomposition of the two-point functions of interest, and explain how they cannot be reproduced by purely unitary representations.

\vskip4pt
\noindent
{\bf Conventions:} Throughout we will work in Euclidean signature for simplicity, despite the fact that we are primarily interested in Lorentzian physics.  Our curvature conventions are those of \cite{Carroll:2004st}. We denote the spacetime dimension by $D$, and denote symmetrization of indices by $(\cdots)$ and anti-symmetrization by $[\cdots]$. We denote the fully traceless symmetrization by $(\cdots)_T$ and all indices are (anti)symmetrized with weight $1$.  In many places we employ Young diagrams and Young tableaux to represent the index symmetries of various tensors. We denote Young diagrams by $[r_1,r_2,\ldots, r_n]$, where $r_a$ are the lengths of the various rows. 
Our conventions are such that all Young diagrams are first symmetrized and then antisymmetrized with respect to the relevant indices (so that all tableaux are in the manifestly antisymmetric presentation). We also define the Hodge star $*$ such that it dualizes the first column of a given tableau.
We denote Young projectors onto the diagram $[r_1,r_2,\ldots, r_n]$ by ${\cal Y}_{[r_1,r_2,\ldots, r_n]}$, and we denote projectors onto the space of traceless tensors by ${\cal Y}^T$. Note that essentially all tensors in this paper are traceless. Typically there is only one  nonzero way to assign indices to be compatible with a given Young projector, so when there is no ambiguity we do not explicitly display the indices involved in the projector.

\section{Symmetries and gapless phases}
\label{sec:general}

In this section we describe the general structure of the relativistic gapless phases of interest, before considering some concrete examples in Sections~\ref{sec:scalarsec} and~\ref{sec:gravsec}. The physics of these phases is controlled by a pair of generalized conserved currents that have a particular mixed anomaly between them. 
The presence of gapless modes is mandated by this anomaly, and we will see through the examples that in some cases the anomaly cannot be reproduced in any unitary theory.

\subsection{Currents and cohomology}

A phase will be defined by specifying both a generalized ``magnetic" current, $K$, and a generalized ``electric" current, $J$, along with the generalized conservation conditions that they each satisfy.  The Lorentz representation of these currents and their conservation conditions do not need to be the same, and will generically not be.  In general, there will be a mixed anomaly between the two symmetries, whose precise form can also be thought of as an input.  A signal of the anomaly will be the impossibility of sourcing one of these currents via a background gauge field without violating conservation of the other current. 
As is usually the case with anomalies, the failure of conservation can be shifted between the currents by adjusting local counterterms in the theory.  As a matter of convention, we will choose the anomaly to appear in the magnetic current's conservation equation.  
 We will see that this mixed anomaly governs the structure of the phase.  At the practical level, we will specify the properties of the gauge field that the current $J$ couples to, rather than the current itself, and this indirectly defines the conservation equation $J$ satisfies. This will both allow us to define the conservation equation for the current $J$, and the gauge-invariant field strength that will ultimately appear in the anomaly equation for $K$.

\subsubsection{Generalized cochain complexes}

The currents can transform in arbitrary representations of the Lorentz group, and satisfy generalized conservation laws~\cite{Pano:2023slc}. In order to describe the systematics, it is convenient to introduce a  cochain complex that generalizes the familiar de Rham complex for antisymmetric tensors to more general representations~\cite{olver1982differential,Dubois-Violette:1999iqe,Dubois-Violette:2001wjr,Bekaert:2002dt} (also known as BGG complexes \cite{92c50ea1-4549-3a51-8381-76d7c6eb773e,eastwood1999variations,Arnold2020ComplexesFC}).
Each of the currents, or its background gauge field, will be an element of one of these complexes, and its conservation law will be described by a generalized exterior derivative that maps between successive elements of the complex. Under the assumption that the cohomology of this complex is trivial, as it is in flat space, we will be able to understand the features of the mixed anomaly of the currents' conservation, the structure of the gauge invariances of external sources, and the structure of the low energy EFT.

It is simplest to consider the elements of the complex as fully traceless tensors, which therefore carry irreducible SO$(D)$ representations in generic $D$. We label the elements by their index symmetries, so that members of the $a^{\rm th}$ element of the complex will have the index symmetries of the Young diagram ${\mathbb Y}^{(a)}$. We can therefore write the cochain complex of interest schematically as
\be
{\mathbb Y}^{(1)}\,
\xrightarrow{\,\rd_{(1)}\,} 
\,{\mathbb Y}^{(2)}\,
\xrightarrow{\,\rd_{(2)}\,}
\,{\mathbb Y}^{(3)}\,
\xrightarrow{\,\rd_{(3)}\,}
\,{\mathbb Y}^{(4)}\,
\xrightarrow{\,\rd_{(4)}\,}
\cdots\,\ \ .
\label{eq:schematiccochain}
\ee
The differential maps $\rd_{(a)}$ will be described shortly; they will map between elements of the complex in such a way that $\rd_{(a+1)}\rd_{(a)}=0$.

We now describe the symmetry types ${\mathbb Y}^{(a)}$ of the elements that can appear in the complex. 
The starting point can be chosen freely. It is an arbitrary tensor with index symmetries given by the Young tableau~${\mathbb Y}^{(1)}$: 
\be
\Yboxdim{13.5pt} 
{\mathbb Y}^{(1)}~\in~\raisebox{18.5pt}{
{\vrule width 1pt}\hspace{-4.5pt}
\gyoung(_6{s_1},_5{s_2},/1\vdts,_3{s_p})
}\,,
\ee
where the integers $s_1,\ldots, s_p$ label the lengths of the rows of the tableau.  The only requirement is that these lengths are non-increasing: $s_1\geq s_2\geq\cdots \geq s_p$, so that ${\mathbb Y}^{(1)}\in [s_1,\ldots, s_p]$. In order to obtain the subsequent elements of the complex~\eqref{eq:schematiccochain}, we add more boxes to the tableaux row-by-row using the integers $r_a$ that we will describe below: to get ${\mathbb Y}^{(2)}$, we add $r_1$ boxes to the first row of ${\mathbb Y}^{(1)}$, to get ${\mathbb Y}^{(3)}$, we add $r_2$ boxes to the second row of ${\mathbb Y}^{(2)}$, and so on.  So in general, to obtain ${\mathbb Y}^{(a+1)}$, we add $r_a$ boxes to the $a$-th row of ${\mathbb Y}^{(a)}$.

The data that we need to completely specify all of the ${\mathbb Y}^{(a)}$ is the first element of the cochain ${\mathbb Y}^{(1)}$, and the number $r_1> 0$ 
of boxes that are added to the first row of ${\mathbb Y}^{(1)}$ in order to obtain ${\mathbb Y}^{(2)}$.  In order to obtain $r_2$, $r_3$, etc. we use the following formula:
\be 
r_a=s^{}_{a-1}-s^{}_{a}+1,\ \ {\rm for} \ \ \ a=2,3,\cdots\,\ \ ,
\ee
where $s^{}_a$ is the number of boxes in the $a$-th row of the initial tableau ${\mathbb Y}^{(1)}$. In particular, this implies that if the starting tableau has $p$ non-zero rows, then ${\mathbb Y}^{(p+3)}$ will have a last row with just a single box, and all the subsequent steps of the algorithm will just add a new one-box row at the bottom of the diagram.  We can visualize the complex as
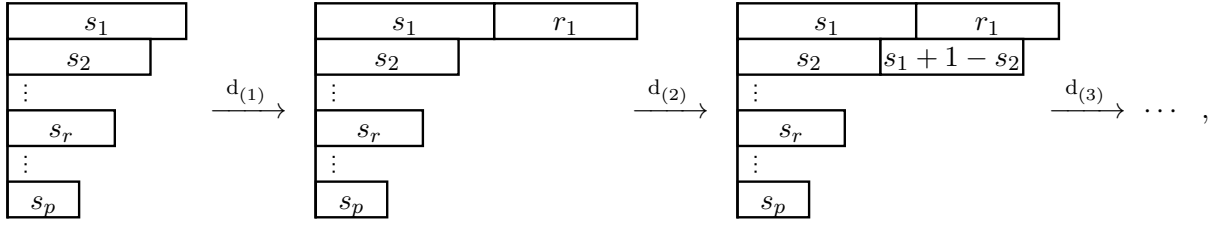
\begin{equation*}
\Yboxdim{13.5pt} 
\raisebox{33pt}{
{\vrule width 1pt}\hspace{-4.5pt}
\gyoung(_5{s_1},_4{s_2},/1\vdts,_3{s_r},/1\vdts,_2{s_p})
}
\,
\xrightarrow{~\rd_{(1)}~}
\,\,
\raisebox{33pt}{
{\vrule width 1pt}\hspace{-4.5pt}
\gyoung(_5{s_1}_4{r_1},_4{s_2},/1\vdts,_3{s_r},/1\vdts,_2{s_p} )
}
\,
\!\!\!\!\!\!
\xrightarrow{~\rd_{(2)}~}
\,\,
\raisebox{33pt}{
{\vrule width 1pt}\hspace{-4.5pt}
\gyoung(_5{s_1}_4{r_1},_4{s_2}_4{s_1+1-s_2},/1\vdts,_3{s_r},/1\vdts,_2{s_p} )
}
\!\!\!\!\!\!
\xrightarrow{~\rd_{(3)}~}
\,\, \cdots \ \ ,
\end{equation*}
which can equivalently be written in terms of the row lengths as
\be
\begin{aligned}
\begin{tikzcd}
\left[ s_1,s_2,s_3,\cdots, s_p\right] \rar{~\rd_{(1)}~} &\left[ s_1+r_1 , s_2,s_3,\cdots, s_p\right] \rar{~\rd_{(2)}~} &
                                                  \left[ s_1+r_1 , s_1+1,s_3,\cdots, s_p\right] 
                                                  \ar[draw=none]{d}[name=X, anchor=center]{}
       & \ar[rounded corners,
            to path={++(-.95,0) -- ([xshift=-4ex]\tikztostart.east)
                      |- (X.center) \tikztonodes
                      -| ([xshift=-2ex]\tikztotarget.west)
                      -- (\tikztotarget)}]{dll}[at end,above]{\hspace{-6.5cm}\rd_{(3)}~} \\[12pt]
{}  &\left[ s_1+r_1 , s_1+1,s_2+1 ,\ldots, s_p\right]  \rar{\rd_{(4)}} & \cdots \ \  .&{} 
\end{tikzcd}
\end{aligned}
\label{eq:complexpicture}
\ee

We now turn to describing the differential maps $\rd_{(a)}$  that take us
between successive element of the complex.   The operator $\rd_{(a)}$ contains $r_a$ derivatives and  maps tensors of type ${\mathbb Y}^{(a)}$ into tensors of type ${\mathbb Y}^{(a+1)}$: operationally, we obtain ${\mathbb Y}^{(a+1)}$ by associating derivatives to the $r_a$ new boxes that are present in ${\mathbb Y}^{(a+1)}$ but not in ${\mathbb Y}^{(a)}$, putting ${\mathbb Y}^{(a)}$ into the remaining boxes and then Young projecting to ${\mathbb Y}^{(a+1)}$ and removing all traces.
We can think of the additional boxes present in the Young diagram ${\mathbb Y}^{(a+1)}$ compared to  ${\mathbb Y}^{(a)}$ as being the derivatives in $\rd_{(a)}$.
Due to the fact that derivatives in successive steps are being added to different rows, and the fact that indices in the same column are antisymmetrized, it is straightforward to see that these operators satisfy the central condition of a cochain complex, that the composition of two successive maps vanishes,
\be 
\rd_{(a+1)}\rd_{(a)}=0,\label{d2eq0e}
\ee
because derivatives end up being antisymmetrized.
By assuming the cohomology of this cochain is trivial, we can say that if a tensor $T^{(a)}$ of symmetry type ${\mathbb Y}^{(a)}$ satisfies $\rd_{(a)}{ T}^{(a)}=0$, then it can be written 
as ${ T}^{(a)}=\rd_{(a-1)} {T }^{(a-1)}$ for some tensor $T^{(a-1)}$ of symmetry type ${\mathbb Y}^{(a-1)}$. In the following, the magnetic conserved current and the gauge field source for the electric current will be elements of complexes of this type.\footnote{Note that these complexes are the same ones underlying the on-shell gauge invariances and Bianchi identities of general mixed-symmetry partially massless fields on (A)dS~\cite{Skvortsov:2006at,Skvortsov:2009zu}.}

\vspace{-4pt}
\begin{oframed}
{\small
\noindent
{\bf\normalsize Examples:} 
Here we give some examples of the above generalized cochain complexes that will be relevant for some of the theories that we study below.
\begin{itemize} 
\item As a familiar first example, consider the starting tensor to be a scalar ${\mathbb Y}^{(1)}=\bullet$, and take $r_1=1$, which gives 
\be
\Yboxdim{10pt} 
\bullet  
\xrightarrow{\,\rd_{(1)}\,} \,
\gyoung(~) 
\xrightarrow{\,\rd_{(2)}\,}
\raisebox{4pt}{
\gyoung(~,~)  
}
\xrightarrow{\,\rd_{(3)}\,}
\raisebox{9.5pt}{
\gyoung(~,~,~)
}
\xrightarrow{\,\rd_{(4)}\,}
\cdots\,.
\ee
This is the standard de Rham complex for $p$-forms, and plays a role in the construction of ordinary superfluids, reviewed in Section~\ref{sec:superfluid}.

\item Another useful complex that will appear in several places takes the same starting point ${\mathbb Y}^{(1)}=\bullet$, but adds two indices in the first step ($r_1=2$), which leads to
\be
\Yboxdim{10pt} 
\bullet  
\xrightarrow{\,\rd_{(1)}\,} \,
\gyoung(~;~) 
\xrightarrow{\,\rd_{(2)}\,}
\raisebox{4pt}{
\gyoung(~;~,~)  
}
\xrightarrow{\,\rd_{(3)}\,}
\raisebox{9.5pt}{
\gyoung(~;~,~,~)
}
\xrightarrow{\,\rd_{(4)}\,}
\cdots\,.
\ee
This complex will be relevant when we study a scalar analogue of conformal gravity in Section~\ref{sec:confscal}, and also appears in the construction of a galileonic superfluidin Section~\ref{sec:galSF}. This is also the cochain complex underlying the gauge invariances of the partially massless graviton~\cite{Hinterbichler:2014xga,Hinterbichler:2016fgl}.\

\item If we take ${\mathbb Y}^{(1)}=\left[1\right]$, and $r_1=1$, we find
\be
\Yboxdim{10pt} 
\gyoung(~) 
\xrightarrow{\,\rd_{(1)}\,} \,
\gyoung(~;~) 
\xrightarrow{\,\rd_{(2)}\,}
\raisebox{4pt}{
\gyoung(~;~,~;~)  
}
\xrightarrow{\,\rd_{(3)}\,}
\raisebox{9.5pt}{
\gyoung(~;~,~;~,~)
}
\xrightarrow{\,\rd_{(4)}\,}
\cdots\,.
\ee
This is the chain complex underlying the diffeomorphism invariance of the linearized massless graviton, and the conserved magnetic current of linearized gravity fits into this complex, as we review in Section~\ref{sec:GR}.
 It will also appear in the construction of the galileon superfluid (Section~\ref{sec:confscal}) and in both the magnetic current and gauge field of the conformal graviton in Section~\ref{sec:confgrav}.

\item As a final example, we can take ${\mathbb Y}^{(1)}=\left[1,1\right]$ and $r_1=1$. 
The corresponding complex is then
\be
\Yboxdim{10pt} 
\raisebox{4pt}{
\gyoung(~,~)
} 
\xrightarrow{\,\rd_{(1)}\,} \,
\raisebox{4pt}{
\gyoung(~;~,~) 
}
\xrightarrow{\,\rd_{(2)}\,} \,
\raisebox{4pt}{
\gyoung(~;~,~;~)  
}
\xrightarrow{\,\rd_{(3)}\,}
\raisebox{9.5pt}{
\gyoung(~;~,~;~,~;~)  
}
\xrightarrow{\,\rd_{(4)}\,}
\raisebox{14pt}{
\gyoung(~;~,~;~,~;~,~)  
}
\xrightarrow{\,\rd_{(5)}\,}
\cdots\,.
\ee
This complex is relevant for the electric gauge field of linearized Einstein gravity, as we will see in Section~\ref{sec:GR}.

\end{itemize}
}
\end{oframed}

\subsubsection{Electric and magnetic currents}

We now want to describe how to use this complex to construct the electric and magnetic currents that we will use to define the phases of interest.
Two objects will play a central role in the construction: a (magnetic) current, $K$, and a gauge field, $A$ which sources the electric current $J$, and so defines it indirectly.  $K$ and $A$ each fit into a complex of the form~\eqref{eq:complexpicture}.  For each, the choice of complex and the position into which it fits is equivalent to choosing the Lorentz representation and conservation equations for the currents.

\vskip2pt
\noindent
{\bf Magnetic current:}
We begin by considering the magnetic current $K$. There are two features that we have to specify: its index symmetries, and the conservation condition that it satisfies. 
We specify its index symmetries via the Young diagram
${\mathbb Y}^{(K)}$ and we take it to be completely traceless.
 In order to specify its conservation conditions, we specify the complex~\eqref{eq:complexpicture} that the magnetic current $K$ belongs to:
\be
\cdots\xrightarrow{\hphantom{\,\rd_{(F)}\,}} {\mathbb Y}^{(\phi)}\,
\xrightarrow{~\rd_{(\phi)}~} 
\,{\mathbb Y}^{(K)}\,
\xrightarrow{\,\rd_{(K)}\,}
\,{\mathbb Y}^{({\cal C})}\,
\xrightarrow{\,\rd_{({\cal C})}\,}
\cdots\,.
\label{eq:Kcomplex}
\ee
Here $K$ is the current of interest, $\phi$ will end up being the field theory degree of freedom that realizes the relevant pattern of symmetries when we construct an EFT, and ${\cal C}$ is a tensor that will appear in the anomalous conservation equation of the magnetic current in the presence of a background gauge field for $J$. {We will see that typically $K$ corresponds to ${\mathbb Y}^{(2)}$ in~\eqref{eq:schematiccochain}---or ${\mathbb Y}^{(3)}$ in cases involving irreducible gauge fields---but in general, we can take $K$ to be a member of any complex that contains an element with the right index symmetries.} (In general, if the current is in the representation ${\mathbb Y}^{(r)}$ then we'll get an EFT involving a gauge field with $r-3$ levels of reducibility.)

The differential operators in the complex are essentially generalized curls or Bianchi identities, so specifying the conservation of $K$ by the differential complex that it fits into may seem somewhat abnormal (usually we think of conserved currents as satisfying divergence-like conditions), but it is actually more natural from the point of view of generating conserved quantities/symmetry operators. Recall that in the familiar case of a conserved current (which satisfies $\partial_\mu J^\mu = 0$), we actually use the dual current $K= \ast J$ (which satisfies $\rd K = 0$) to define a conserved quantity by integrating over a codimension-1 surface. Here we are specifying this current that satisfies an exterior derivative-like conservation condition directly.\footnote{Note that this is a slightly different perspective/convention from that  employed in~\cite{Hinterbichler:2022agn}, where the magnetic current called $K$ here would be called $*K*$ there in the generic case.} Of course one could translate these conditions into divergence-like conditions on the Hodge dual of $K$, fitting into a dual complex, if desired.

\vskip2pt
\noindent
{\bf Electric gauge field:} Next we specify the background gauge field $A$ that couples to the conserved ``electric'' current $J$. We will take $A$ to be traceless with the index symmetries of a tableau ${\mathbb Y}^{(A)}$.  The current then has the same index symmetries: $J \in {\mathbb Y}^{(A)}$. Like the magnetic current, we then have to specify what complex of the form~\eqref{eq:complexpicture} the gauge field $A$ is a member of. In this case, the complex determines the gauge transformation rule for $A$:
\be
\cdots\xrightarrow{\hphantom{\,\rd_{(F)}\,}} {\mathbb Y}^{(\Lambda)}\,
\xrightarrow{~\rd_{(\Lambda)}~} 
\,{\mathbb Y}^{(A)}\,
\xrightarrow{~\rd_{(A)}~}
\,{\mathbb Y}^{(F)}\,
\xrightarrow{~\rd_{(F)}~}
\cdots\,.
\label{eq:acomplex}
\ee
In this complex, $A$ is the background gauge field, $\Lambda$ is a gauge parameter with the index symmetries of the tableau ${\mathbb Y}^{(\Lambda)}$, so that
$A$ transforms as
\be
\delta A_{{\mathbb Y}^{(A)}} = \rd_{(\Lambda)}\Lambda_{{\mathbb Y}^{(\Lambda)}}\,,
\label{eq:agaugetranfsec1}
\ee
and $F$ is the field strength with the with the index symmetries of the tableau~${\mathbb Y}^{(F)}$, which is gauge invariant by virtue of \eqref{d2eq0e}, \be \delta F_{{\mathbb Y}^{(F)} }=0\, .\ee 

Specifying the pattern of gauge symmetry for $A$ is equivalent to specifying the conservation condition for $J$: since $A$ couples to the current $J$, the current must satisfy a conservation equation of the form
\be
*\,\rd_{(A)}*J_{{\mathbb Y}^{(A)}} = 0\,,
\label{eq:AgaugeJ}
\ee
in order for the coupling to be gauge invariant.

\subsection{Anomalies and gapless modes}

Fully defining the phases of interest requires one additional piece of information: the systems have two conserved currents, $J$ and $K$, and we have to specify the compatibility of their conservation laws. A priori, it is not guaranteed in the presence of a source for one of these conserved currents that the other will continue to be conserved. Indeed, in many cases conservation will fail, in which case we say there is a mixed 't Hooft  anomaly between these symmetry currents. By specifying a particular structure of this anomaly, we will see that the system must be in a gapless phase.

In the presence of the background gauge field $A$ (which sources $J$), both $K$ and $J$ are promoted to gauge-invariant versions ${\cal J}$ and ${\cal K}$. 
Conservation of the current ${\cal J}$ can be maintained by adding appropriate local counterterms,
 but the magnetic current ${\cal K}$ will generically not be conserved. It will instead satisfy the anomalous conservation equation\footnote{The $-$ sign is conventional, and we normalize the anomaly so that properly quantized operators charged under $K$ have unit charge.}
\be
\rd_{(K)}{\cal K}_{{\mathbb Y}^{(K)}}  = -{\cal C}_{{\mathbb Y}^{({\cal C})}}\,,
\label{eq:Kanomeq}
\ee
where ${\cal C}$ is a tensor built from the gauge-invariant field strength $F$ appearing in~\eqref{eq:acomplex} and which has the index symmetries ${\mathbb Y}^{({\cal C})}$ of the tensor appearing in~\eqref{eq:Kcomplex}.\footnote{In all the cases we consider, the anomaly can be written in terms of a gauge-invariant field strength. This is because the anomalies that we consider are ``Abelian". For non-Abelian anomalies, it is typically not possible to write the anomaly in terms of a gauge-invariant combination that also satisfies the Wess--Zumino consistency conditions~\cite{Bardeen:1984pm}.} In many cases $F$ and ${\cal C}$ do not have the same symmetry type. If this happens, since $F$ is traceless, the only possible way to construct a tensor with the correct symmetry properties is by taking derivatives of $F$.

The failure of conservation encapsulated by~\eqref{eq:Kanomeq} is the sign of a mixed anomaly between the electric and magnetic symmetries. The structure of the phase is essentially dictated by this anomaly. One way to understand this is to consider the two-point function between the currents $J$ and $K$. It is fixed by the requirements that $J$ and $K$ be conserved at separated points, and by the anomaly equation~\eqref{eq:Kanomeq}, which specifies the contact terms that spoil conservation of the current $K$ at coincident points. (The fact that it is impossible to have both currents conserved even at coincident points is another way of expressing the anomaly.) 
The spectral decomposition of the two point function $\langle JK\rangle$ then provides us with information about the degrees of freedom in this phase.
 In all cases of interest in this paper, $K$  has the same symmetry type as $J$, and the \KL spectral decomposition of their two-point function will require the presence of gapless modes, whose precise identity depends on the symmetry type and conservation conditions of the currents, along with the anomaly equation.\footnote{This structure of anomalies is what appears in generic spacetime dimension. However, there can be certain special dimensions where interesting additional features appear. In particular, it is sometimes possible to construct gauge-invariant topological terms out of the gauge field $A$, which can generate trace anomalies for the currents of interest (particularly ${\cal J}$). Some examples of these trace anomalies were discussed in~\cite{Hinterbichler:2022agn}. We will not consider them further, but it would be very interesting to understand whether these topological terms contain any interesting information about the global structure of the manifolds on which they are defined.}

Certain gapless phases can therefore be {\it defined} by their structure of conserved currents and anomalies. It would not be possible to match the two-point function of these currents without gapless modes present in the system. We will see in some cases that the two-point function cannot be matched by any spectrum of unitary representations of the Poincar\'e group, implying that the relevant pattern of symmetries cannot arise as the effective description of any unitary quantum system, i.e. it is {\it impossible}.

\subsection{Effective field theory}

It is often desirable to have an effective description of small fluctuations around the ground state of a system in a given phase. We now turn to the construction of such an EFT description of the phases of interest, which realizes the symmetry and anomaly structure. This discussion will be somewhat abstract, but we will  later provide a number of concrete examples.

To build an effective field theory, we need to introduce some local field theory degrees of freedom $\phi$ from which the currents $J$ and $K$ will be built. Since the anomaly~\eqref{eq:Kanomeq} involves the magnetic current, $K$, it is convenient to realize this current as a ``topological" current, whose conservation is trivialized by the parameterization of degrees of freedom.  This is a choice: by choosing appropriate counterterms, we could instead make the anomaly appear in the conservation of $J$, and instead realize $J$ as a topological current in dual variables. This would change the presentation of the EFT, but would not change any of the actual long distance physics of the system. With this choice, the magnetic conserved current will not get deformed in the presence of abelian interactions, so that the anomaly structure is reliable in the interacting theory. Looking at the complex~\eqref{eq:Kcomplex}, we see that if we define
\be
K \equiv \rd_{(\phi)}\phi\,,
\label{eq:phidef}
\ee
then conservation of $K$ will be automatic: $\rd_{(K)}K = \rd_{(K)} \rd_{(\phi)}\phi =0$. 

Reproducing the relevant anomaly in this effective description is fairly straightforward. We promote the magnetic current to its gauge invariant version:
\be
K \to {\cal K} \equiv K - C\,,
\ee
where $C$ is a tensor built out of the gauge field $A$ with the same index symmetries as $K$. This tensor is uniquely defined by the requirement that ${\cal K}$ is gauge invariant. In the presence of $C$,
${\cal K}$ is no longer annihilated by $\rd_{(K)}$, and instead satisfies~\eqref{eq:Kanomeq} with ${\cal C} = \rd_{(K)}C$. Note also that ${\cal C}$ can be written in terms of (possibly derivatives of) the gauge-invariant field strength $F$ associated to $A$.

We now want to construct a Lagrangian involving $\phi$ whose equations of motion imply the conservation of the current $J$.
There are two conceptually different structures that the effective field theory can have. The first is {\it Maxwell-like}, where the effective action is just given by powers of~\eqref{eq:phidef}
\be
S = \int\rd^Dx \left(\frac{1}{2}\left(\rd_{(\phi)}\phi\right)^2+c_1\left(\rd_{(\phi)}\phi\right)^3+\cdots\right)\,.
\label{eq:maxwellEFT}
\ee
In effective field theories of this type, the tensor $C$ is actually just directly the gauge field $A$. Consequently, building the action out of powers of $K$ guarantees that the interacting EFT can be consistently coupled to the gauge field $A$. The corresponding current is $J$, which therefore satisfies the conservation condition~\eqref{eq:AgaugeJ}.\footnote{Notice also that the form of the action guarantees that it will have the shift symmetry
\be
\phi \mapsto \phi+ \alpha_{{\mathbb Y}^{(\phi)}}\,,\label{shiftsymKe}
\ee
where $\alpha$ is a tensor with the same symmetry type as $\phi$, which also satisfies $\rd_{(\phi)}\alpha = 0$. This signals that the symmetry associated with $J$ is spontaneously broken in this phase, and $\phi$ serves as the corresponding Goldstone boson.}

In some cases the equations of motion following from~\eqref{eq:maxwellEFT} involve more than two derivatives. This may be undesirable, and we may want to find theories with the same degrees of freedom and symmetries but with lower-order equations of motion. This can sometimes be achieved by constructing an EFT that is {\it Chern--Simons-like}, i.e., using terms that are not constructed solely from $K$ yet are still invariant under~\eqref{shiftsymKe} up to a total derivative (often known as Wess--Zumino terms).  To our knowledge there is no classification of such terms for totally generic symmetry types (though many scalar cases are known~\cite{Goon:2012dy,Hinterbichler:2014cwa,Griffin:2014bta}), and so we will just cover the cases of later interest.

What we require in these cases is an analogue of the Einstein tensor. This can be constructed by considering the trace-ful analogue of the complex that the current $K$ fits into. We can then define the relevant Einstein tensor as the traceless part of the single trace of $K$ in this complex:
\be
G = {\cal Y}^T_{{\mathbb Y}^{({\rm tr}K)}}\,{\rm tr}\,K\,,
\ee
where $K$ can be written in terms of $\phi$.
In all the cases of interest $\phi$ will have two fewer indices than $K$, and so it is possible to contract $\phi$ with $G$,\footnote{This is the difficulty in describing the totally generic construction for currents of any symmetry type, because $\phi$ and $G$ will not always have the same indices.}
 producing an action of the form
\be
S = \int\rd^Dx \Big( \phi \,G+c_1\left(\rd_{(\phi)}\phi\right)^2+c_2\left(\rd_{(\phi)}\phi\right)^3+\cdots\Big)\,,
\ee
where we have allowed for the possibility of irrelevant corrections built out of $\phi$, which may also be of Wess--Zumino type. It is somewhat non-obvious, but this action will have the same global symmetries as~\eqref{eq:maxwellEFT}, but with the Einstein term being invariant only up to a total derivative. The equation of motion following from this action is a partial flatness condition on the trace-ful version of the curvature $K$, in contrast to the wave equation-like condition following from~\eqref{eq:maxwellEFT}.

We will see versions of both of these types of  EFT in the following examples. Heuristically, one can think of the Maxwell-like EFTs as being more like electromagnetism, and the Chern--Simons-like EFTs as being more like Einstein gravity. While the precise EFT construction may seem like somewhat of a choice at the level of global symmetries, there actually is no freedom once one specifies the anomaly structure that the EFT is to reproduce. One way to understand this is that in the deep IR, the two currents $J$ and $K$ coincide, and so the two-point function of these currents is determined by the free theory, and the number of derivatives in the quadratic action will affect the form of this two-point function. Since the two-point function of the theory is completely determined by the conservation conditions and anomalies, different quadratic actions must have different anomaly structures. Thus, the anomaly that we are trying to reproduce dictates the kinetic structure that must be present.
In Section~\ref{sec:scalarsec} we will see an explicit example, where the current $K$ will be the same in two different theories, but the current $J$ will satisfy different conservation conditions depending on the kinetic term of the theory, and correspondingly the theories can be coupled to different background gauge fields and will produce different anomalies.

\section{Scalar examples}
\label{sec:scalarsec}

We now turn to some concrete examples of the preceding abstract discussion.  In this section we present the simplest possible cases: a scalar field realizing various patterns of symmetries and anomalies which fit into the general framework presented in Section~\ref{sec:general}.

\subsection{Superfluid}
\label{sec:superfluid}

One of the simplest examples of a gapless quantum system is a superfluid. The massless phonon in this phase can be understood as a consequence of a mixed anomaly between the shift symmetry acting on the phonon and an (emergent) winding symmetry, under which the superfluid's vortices are charged~\cite{Delacretaz:2019brr}. Here we review this in the context of the general construction of Section~\ref{sec:general}.

\vskip2pt
\noindent
{\bf Currents and anomalies:} In the language of Section~\ref{sec:general}, the two objects of interest are a magnetic $1$-form current $K_{\mu}$ and a $1$-form gauge field $A_\mu$, which sources an electric current $J_\mu$. Both the current $K_\mu$ and gauge field $A_\mu$ can be thought of as elements of the ordinary de Rham complex
\be
\Yboxdim{12pt} 
\bullet  
\xrightarrow{~\,\rd\,~} \,
\gyoung(~) 
\xrightarrow{~\,\rd\,~}
\raisebox{5pt}{
\gyoung(~,~)  
}
\xrightarrow{\,~\rd~\,}
\cdots\,.
\label{eq:superfluidderham}
\ee
Here, the presence of $K_\mu$ in this complex signals that it is closed $\partial_{[\mu}K_{\nu]} = 0$, which implies that its dual $(*K)_{\mu_1\cdots\mu_{D-1}}$ is conserved in the ordinary sense, $\partial^{\mu_1}(*K)_{\mu_1\cdots\mu_{D-1}}=0$.  The fact that $A_\mu$ is part of this complex implies that it has the usual gauge transformation rule $\delta A_\mu = \partial_\mu\Lambda$, and thus sources a current $J_\mu$ which is conserved in the ordinary sense, $\partial_\mu J^\mu=0$.
These equations are true as operator statements at separated points, but they cannot  both be chosen to be true at coincident points in correlation functions. Contact terms necessarily spoil conservation of at least one of them. The structure of these contact terms is dictated by the anomaly equation\footnote{As mentioned above, we have chosen to put the anomaly in the conservation equation for $K_\mu$. We could instead choose a renormalization scheme that shuffles the anomaly into the conservation equation for $J$. This other choice would be more natural in dual variables where the superfluid phonon is packaged into a $(D-2)$-form (see e.g.,~\cite{Horn:2015zna}).}
\be
\partial_{[\mu}{\cal K}_{\nu]} = -\frac{1}{2}F_{\mu\nu}\,,
\label{eq:sfluidanom}
\ee
where $F_{\mu\nu} = \partial_\mu A_\nu - \partial_\nu A_\mu$ is the field strength that is part of the complex~\eqref{eq:superfluidderham}, and ${\cal K}_\mu$ is the gauge-invariant version of $K$ in the presence of $A_\mu$. 
So we see here that the gauge-invariant field strength directly plays the role of ${\cal C}$ in~\eqref{eq:Kanomeq}.

In order to see how this pattern of symmetries and anomalies defines the superfluid phase, we note that the Fourier-space two-point function between $J$ and $K$ is completely fixed by their conservation and mixed anomaly \eqref{eq:sfluidanom} to be~\cite{Delacretaz:2019brr}
\be
\langle J_\mu K_\nu\rangle = \frac{1}{p^2}\left(p_\mu p_\nu-p^2\eta_{\mu\nu} \right)\,.
\ee
The pole at $p^2 \to 0$ indicates the presence of a gapless excitation in the spectrum, and a more careful spectral decomposition reveals that this excitation is a scalar particle~\cite{Delacretaz:2019brr,Hinterbichler:2022agn}. From this perspective, the superfluid Goldstone mode is forced upon us by the currents and the structure of their anomaly.

\vskip2pt
\noindent
{\bf Effective field theory:} It is straightforward to construct an EFT that reproduces this physics and describes the superfluid phonon directly~\cite{Son:2002zn}. Recall that the (magnetic) current satisfies $\rd K = 0$. From the complex~\eqref{eq:superfluidderham} we see that we can trivialize this conservation law by writing
\be
K_\mu = \partial_\mu\phi\,,\label{eq:Ksuperfluid}
\ee 
for some scalar field $\phi$. Our goal is then to construct a field theory for $\phi$ that has a conserved (electric) current. In this case the relevant EFT is Maxwell-like, and we can write
\be
S = \int\rd^Dx\left(\frac{1}{2}(\partial\phi)^2+\frac{1}{4\Lambda^{D}}(\partial\phi)^4+\cdots
\right)\,,
\label{eq:sfluidEFT}
\ee
where $\Lambda$ is a dimension-ful scale.
The existence of a conserved electric current that is conserved on-shell follows from the shift symmetry $\phi\mapsto\phi+c$  of the action,
\be
J_\mu = \partial_\mu\phi\left[1 +\frac{1}{\Lambda^D}(\partial\phi)^2+\cdots\right]\,.
\ee
 We can couple this current to a background gauge field $A_\mu$ by promoting everywhere
\be
\partial_\mu\phi ~\mapsto~ \partial_\mu\phi-A_\mu\,,
\ee
after which the action is invariant under the gauge transformations $\delta\phi= \xi(x)\,,\delta A_\mu = \partial_\mu \xi(x)$. In this case, the magnetic current \eqref{eq:Ksuperfluid} fails to be gauge invariant, and we then see that the gauge invariant version of this current
\be
{\cal K}_\mu \equiv \partial_\mu \phi - A_\mu\,,
\ee
no longer is anti-symmetrically conserved, but instead displays the expected anomaly~\eqref{eq:sfluidanom}. We therefore see that the EFT~\eqref{eq:sfluidEFT} captures the low-lying excitations of this phase, with $\phi$ acting as the superfluid phonon. 

\vskip2pt
\noindent
{\bf Two dimensions:} A small interesting feature appears in the linearized theory in two dimensions.
 In this case, the linear equations of motion are
\be
\partial_\mu J^{\mu} = 0\,,\hspace{2cm} \partial_{\mu_1}(*J)^{\mu_1\cdots\mu_{D-1}}=0\,.
\ee
so we see that precisely in $D=2$ these are electric-magnetic duals of each other.

\subsection{Galileonic superfluid}
\label{sec:galSF}

As another example with the same degrees of freedom as an ordinary superfluid---but with different symmetries---we now consider a ``galileon superfluid,'' which is an exotic version of a superfluid with a spacetime-dependent shift symmetry~\cite{Nicolis:2008in,Hinterbichler:2022agn}.

\vskip2pt
\noindent
{\bf Currents and anomalies:} As before, we begin by introducing the magnetic current and the electric gauge field, which will be used to define the phase. In this case, the magnetic current is a traceless symmetric two-index tensor
\be
K_{\mu\nu}
~\in~
\Yboxdim{12pt} 
\gyoung(~;~) \,\,.
\ee
We consider $K$ to be part of the complex
\be
\Yboxdim{12pt} 
\bullet  
\xrightarrow{\,\rd_{(\phi)}\,} \,
\gyoung(~;~) 
\xrightarrow{\,\rd_{(K)}\,}
\raisebox{5pt}{
\gyoung(~;~,~)  
}
\xrightarrow{\,\hphantom{\rd_{(3)}}\,}
\cdots\,,
\label{eq:galsfKcompl}
\ee
where the maps appearing are
\begin{align}
\label{eq:phimap}
(\rd_{(\phi)}\phi)_{\mu\nu} &= \partial_{(\mu}\partial_{\nu)_T}\phi\,,\\
(\rd_{(K)}K)_{\alpha\mu\nu} &= 3{\cal Y}^T_{[2,1] }\partial_{\alpha}K_{\mu\nu} = \partial_\alpha K_{\mu\nu}+\cdots\,,
\label{eq:gsfKcons}
\end{align}
so that the conservation condition satisfied by $K$ is $\rd_{(K)}K= 0$, which involves  projecting onto the traceless $[2,1]$ tableau.

We specify the electric current by defining a symmetric traceless gauge field $A_{\mu\nu}$, which is an element of the complex,
\be
\Yboxdim{12pt} 
\gyoung(~) 
\xrightarrow{\,\rd_{(\Lambda)}\,} \,
\gyoung(~;~) 
\xrightarrow{\,\rd_{(A)}\,}
\raisebox{5pt}{
\gyoung(~;~,~;~)  
}
\xrightarrow{\,\rd_{(F)}\,}
\cdots\,\ \ .
\ee
This implies that the gauge transformation rule for $A$ is $\rd_{(\Lambda)}\Lambda$ which can be written as
\be
\delta A_{\mu\nu} = 2\partial_{(\mu}\Lambda_{\nu)_T}\,,
\label{eq:gsfAgauge}
\ee
and the gauge-invariant field strength $F = \rd_{(A)}A$ has the symmetries of the Weyl tensor
\be
F_{\mu\nu\alpha\beta} = \frac{3(D-1)}{2(D-3)} {\cal Y}^T_{[2,2]} \,\,\partial_{\mu}\partial_{\alpha}A_{\nu\beta}=\frac{(D-1)}{(D-3)}\partial_{\mu}\partial_{\alpha}A_{\nu\beta}+\cdots\,,
\label{eq:gsffieldstrength}
\ee
where we have chosen the normalization for later convenience.
The gauge field $A$ sources the electric current $J_{\mu\nu}$.  The current $J_{\mu\nu}$ therefore has the same symmetry properties (symmetric traceless) and the gauge transformation~\eqref{eq:gsfAgauge} implies that it satisfies the conservation condition
\be \partial^\nu J_{\mu\nu} = 0.\ee

In the presence of the background gauge field $A$, the gauge-improved current ${\cal J}$ continues to be conserved, but the magnetic current no longer satisfies~\eqref{eq:gsfKcons}, but rather obeys the anomalous conservation equation
\be
{\cal Y}^T_{[2,1]}\,\partial_{\alpha}{\cal K}_{\mu\nu} = - {\cal C}_{\alpha\mu\nu}\, ,
\label{eq:galanom}
\ee
where the right-hand side is built from the field strength~\eqref{eq:gsffieldstrength} as 
\be
{\cal C}_{\alpha\mu\nu} =- \partial^\beta F_{\beta \nu\alpha\mu} =  \frac{3}{2}{\cal Y}^T_{[2,1]} \partial_{\alpha}\, C_{\mu\nu}\, ,\label{CtoFeqe}
\ee
where 
\be  C_{\mu\nu} =   \frac{D-1}{D-2}\bigg[2\partial^\alpha \partial_{(\mu}A_{\nu)_T \alpha}-\square A_{\mu\nu}\bigg]\,.
\label{eq:galCtens}
\ee
Notice that the tensor appearing in the anomaly ${\cal C}_{\alpha\mu\nu}$ is not directly $F_{\mu\nu\alpha\beta}$ because of the mismatch in index structures: since ${\cal C}$ has one fewer index than $F$, it must be constructed using one derivative, and \eqref{CtoFeqe} is the unique way of making something with the index symmetries of ${\cal C}$ out of a derivative of $F$.
Similarly to the ordinary superfluid, this anomaly fixes the two-point function between the currents $J_{\mu\nu}$ and $K_{\mu\nu}$ to be
\be
\begin{aligned}
\braket{{J}_{{\mu_1}{\mu_2}}{K}_{{\nu_1}{\nu_2}}} = \frac{D-1}{D-2}\Bigg[p^2\eta_{{\mu_1}({\nu_2}}\eta_{{\nu_1}){\mu_2}} &- \frac{1}{D-1}\Big(p^2 \eta_{{\mu_1}{\mu_2}}\eta_{{\nu_1}{\nu_2}} -\eta_{{\mu_1}{\mu_2}}p_{\nu_1} p_{\nu_2}-\eta_{{\nu_1}{\nu_2}}p_{\mu_1} p_{\mu_2} \Big)\\
& \hspace{-5pt}- \eta_{{\mu_2}({\nu_2}}p_{{\nu_1})}p_{\mu_1} -\eta_{{\mu_1}({\nu_2}}p_{{\nu_1})}p_{\mu_2}+ \frac{D-2}{D-1}\frac{p_{\mu_1} p_{\mu_2} p_{\nu_1} p_{\nu_2}}{p^2}\Bigg] \,,\label{eq:CurrCurre2pt}
\end{aligned} 
\vphantom{\Bigg\}}
\ee
and the \KL spectral decomposition once again reveals that the gapless excitation in the system is a Poincar\'e scalar~\cite{Hinterbichler:2022agn}.

In $D=2$ there is a topological term that can be built out of the gauge field $A_{\mu\nu}$ which has no free indices (it is the linearized Ricci tensor $\partial^\mu\partial^\nu A_{\mu\nu}$, thinking of $A_{\mu\nu}$ as a traceless version of the graviton). This topological term can appear in the trace equations $\tr {\cal J} = \tr {\cal K} = 0$, leading to additional anomalies~\cite{Hinterbichler:2022agn}.

\vskip2pt
\noindent
{\bf Effective field theory:} We would now like to construct an EFT that reproduces the anomaly structure and describes this phase. From the complex~\eqref{eq:galsfKcompl}, we see that we can trivialize the magnetic conservation equation by defining
\be
K_{\mu\nu} = \partial_{(\mu}\partial_{\nu)_T}\phi = \partial_\mu\partial_\nu\phi-\frac{1}{D}\eta_{\mu\nu}\square\phi\,.
\ee
Recall that for this current, the desired conservation equation is that the antisymmetric derivative of $K_{\mu\nu}$ with all traces removed vanishes, as dictated by the complex~\eqref{eq:galsfKcompl}, and this is guaranteed by writing $K$ in this way.
As expected, the relevant degree of freedom in this phase is a scalar field $\phi$. We now want to construct an action that leads to the conservation of the electric current $J_{\mu\nu}$. Notice that $K_{\mu\nu}$ is invariant under the transformation~\cite{Nicolis:2008in}
\be
\delta \phi = c_\mu x^\mu\,,
\label{eq:galsymm}
\ee
and so it is natural to imagine that $J_{\mu\nu}$ will be the Noether current for this symmetry.\footnote{Notice that this EFT is still written in terms of a scalar field, despite the fact that it has {\it more} symmetry than the superfluid (it is also clearly invariant under $\delta\phi={\rm const.}$). The fact that a single degree of freedom can realize multiple symmetries is a manifestation of the so-called inverse Higgs effect~\cite{Ivanov:1975zq}.} 
More specifically, we want to construct an EFT that has a global symmetry $\delta\phi = \partial_\mu \xi^\mu$, where $\xi^\mu$ is a conformal Killing vector that also satisfies $\partial_\mu\partial_\nu\partial_\alpha\xi^\beta = 0$~\cite{Hinterbichler:2022agn}.
One way to guarantee that this will be a symmetry of the action is to construct the action out of powers of $K_{\mu\nu}$, which would be analogous to the Maxwell-like EFT of a superfluid~\eqref{eq:sfluidEFT}. However, this action will necessarily have higher-derivative equations of motion, and it would not reproduce the anomaly~\eqref{eq:galanom}. We therefore are motivated to ask whether it is possible to construct an EFT with the same symmetries, but which has second order linearized equations of motion. 
This turns out to indeed be possible. Notice that the linearized action
\be
S = -\int\rd^Dx \,\frac{1}{2}\phi\square\phi\,,
\label{eq:freegal}
\ee
is also invariant under the symmetry~\eqref{eq:galsymm}, despite having fewer derivatives.\footnote{We can construct the ``Einstein scalar" $G = -\square\phi$ from the trace-ful version of the complex~\eqref{eq:galsfKcompl}. In this complex, we would have $K_{\mu\nu} =\partial_\mu\partial_\nu\phi$, which has $G$ as its nonzero trace.}
This is possible because the action is invariant up to a total derivative under the symmetry---there also exist interactions with the same property, so called galileon terms~\cite{Fairlie:1992nb,Nicolis:2008in}. We can gauge the symmetry~\eqref{eq:galsymm} by promoting $\square\phi$ to
\be
{\cal G} =-\square\phi + \partial_\mu \partial_\nu A^{\mu\nu}\,,
\ee
after which the linearized action can be put in the form
\be 
S = \frac{1}{2}\int \rd^D x\Big(\phi \,\mathcal{G}+ A^{\mu\nu}\mathcal{J}_{\mu\nu}\Big)\, .\label{eq:S324}
\ee
Here ${\cal J}_{\mu\nu}$ is the electric current
\be
{\cal J}_{\mu\nu} =  \partial_{(\mu}\partial_{\nu)_T}\phi-C_{\mu\nu}\,,
\label{eq:Jcurrentgauge2}
\ee
where $C_{\mu\nu}$ is given by~\eqref{eq:galCtens}.
In the linearized theory, it happens that ${\cal K}_{\mu\nu} = {\cal J}_{\mu\nu}$, but this will not be true generically once interactions are included. Taking the antisymmetric derivative, we reproduce~\eqref{eq:galanom}~\cite{Hinterbichler:2022agn}. The action~\eqref{eq:S324} is invariant under the gauge transformation
\be 
\delta\phi = \frac{2(D-1)}{D}\partial^\mu\xi_\mu\, ,
\ee
for $\phi$, along with the transformation~\eqref{eq:gsfAgauge} for $A$.

Interactions can be introduced similarly by taking arbitrary powers of ${\cal K}_{\mu\nu}$, which coincides with~\eqref{eq:Jcurrentgauge2}. The magnetic current (and anomaly) do not change. The electric current will receive corrections from the interaction terms, but will continue to exist as long as the interactions respect the symmetry~\eqref{eq:galsymm} (or equivalently gauge invariance in the presence of $A$). Note that this theory and the ordinary superfluid have the same kinetic term.  The difference between them lies only in the symmetries that are demanded, and thus in the spectrum of operators and in the types of irrelevant interactions that we are allowed to include.\footnote{Note that if we had not included the $\phi\,{\cal G}$ term and instead built the EFT directly out of powers of ${\cal K}$, we would see that ${\cal K}$ has an anomaly of the form
\be
{\cal Y}^T_{[2,1]}\,\partial_{\alpha}{\cal K}_{\mu\nu} \sim - \square\,{\cal C}_{\alpha\mu\nu}\, ,
\ee
rather than~\eqref{eq:galanom}, and correspondingly this EFT would have a different two-point function.
}

\vskip2pt
\noindent
{\bf Two dimensions:} Here again there is a version of electric-magnetic duality in two dimensions.
This case is similar to the ordinary superfluid, but just with an extra index. The two equations of motion are
\be
\partial_\mu J^{\mu\nu} = 0\,,\hspace{2cm} \partial_{\mu_1}(*J)^{\mu_1\cdots\mu_{D-1}\,\nu}=0\,,
\ee
where $*J$ is dualized over its $\mu$ index. We again see that these equations are duals of each other in $D=2$.

\subsection{Conformal scalar}
\label{sec:confscal}
We now want to consider a more novel example, which is a conformal scalar that shares many similarities with conformal gravity.
Here we meet our first example of a theory with ``impossible" symmetries, where the structure of anomalies cannot be matched by a unitary theory.
Consequently we will find that the EFT that captures the dynamics and reproduces the anomalies necessarily involves higher-order derivatives.

\vskip2pt
\noindent
{\bf Currents and anomalies:} In the previous examples, the currents of interest satisfied single-derivative conservation equations. In this example we will have an electric current that satisfies a higher-derivative condition. 

The magnetic current is the same as in the galileon superfluid example
\be
K_{\mu\nu}
~\in~
\Yboxdim{12pt} 
\gyoung(~;~) \,\,,
\ee
which then fits into the same complex
\be
\Yboxdim{12pt} 
\bullet  
\xrightarrow{\,\rd_{(\phi)}\,} \,
\gyoung(~;~) 
\xrightarrow{\,\rd_{(K)}\,}
\raisebox{5pt}{
\gyoung(~;~,~)  
}
\xrightarrow{\,\hphantom{\rd_{(3)}}\,}
\cdots\, \ \ .
\label{eq:CSsKcomp}
\ee
In this complex, the maps are the same as~\eqref{eq:phimap} and~\eqref{eq:gsfKcons},
and the conservation condition that $K$ satisfies is $\rd_{(K)}K = 0$.

The novelty in this case compared with the superfluid is the electric gauge field: it is once again a two-index traceless tensor $A_{\mu\nu}$, but it is now part of the complex
\be
\Yboxdim{12pt} 
\bullet  
\xrightarrow{\,\rd_{(\Lambda)}\,} \,
\gyoung(~;~) 
\xrightarrow{\,\rd_{(A)}\,}
\raisebox{5pt}{
\gyoung(~;~,~)  
}
\xrightarrow{\,\rd_{(F)}\,}
\cdots\,.
\label{eq:confscalAcomplex}
\ee
Under a gauge transformation, $A$ shifts as
\be
\delta A_{\mu\nu} = \partial_{(\mu}\partial_{\nu)_T}\Lambda\,,
\label{eq:AgaugeCS}
\ee
and the field strength $F_{\mu\nu\alpha}$ is defined as
\be
F_{\mu\nu\alpha} =\partial_{[\mu}A_{\nu]\alpha}-\frac{1}{D-1}\eta_{\alpha[\mu}\partial^\beta A_{\nu]\beta}\,.
\ee
These two equations are the maps in~\eqref{eq:confscalAcomplex} expressed in indices.
The gauge transformation~\eqref{eq:AgaugeCS} implies that the electric current satisfies the conservation condition
\be
\partial_{\mu}\partial_\nu J^{\mu\nu} = 0\,.
\ee
Note that the single divergence of $J$ does not necessarily vanish, $\partial_\nu J^{\mu\nu}\neq 0$.

In the presence of the background gauge field $A_{\mu\nu}$, the electric current ${\cal J}_{\mu\nu}$ continues to be conserved, but the magnetic current instead satisfies the anomalous conservation equation
\be
{\cal Y}_{[2,1]}^T\,\partial_{\alpha}{\cal K}_{\mu\nu} = - F_{\alpha\mu\nu}\,.
\label{eq:confscalaranom}
\ee
Since $F_{\alpha\mu\nu}$ has the correct index symmetries, it can appear directly on the right-hand side of this equation, playing the role of ${\cal C}_{\alpha\mu\nu}$ in~\eqref{eq:Kanomeq}. In this sense the mixed anomaly for this theory is conceptually more similar to that of the ordinary superfluid than the galileon superfluid, despite the fact that the magnetic current has the same index symmetries as in the galileonic example.

As in the previous examples, the conservation condition $\partial_\mu\partial_\nu J^{\mu\nu} = 0$ along with the mixed anomaly~\eqref{eq:confscalaranom} completely fixes the Fourier-space two-point function between the two currents:
\be
\braket{J_{\mu\nu}K_{\rho\sigma}} = \frac{D}{D-1}\frac{1}{p^4}\left[\,p_\mu p_\nu p_\rho p_\sigma-\frac{D-1}{D}p^4 \eta_{\mu(\sigma}\eta_{\rho)\nu}+\frac{p^2}{D} \left(p^2 \eta_{\mu\nu}\eta_{\rho \sigma}- p_{\mu}p_{\nu} \eta_{\rho\sigma}- p_{\rho}p_{\sigma}\eta_{\mu\nu}\right)\right]\, .\label{eq:TwoPFConfScalar}
\ee
Notice that this two-point function has a $p^{-4}$ divergence as $p\to 0$, which already suggests that this theory cannot be unitary. This can be formalized by an explicit spectral decomposition,
but the result is the expected one---the higher-order pole cannot be matched by any unitary Poincar\'e representation, and instead reveals the existence of a so-called ``dipole ghost'' state~\cite{Karowski:1974qx,Binegar:1983kb,Berkovits:2004jj}. More details are provided in Appendix~\ref{ap:Spec}. We therefore see that any theory with this pattern of symmetries and anomalies must necessarily be non-unitary---they are impossible symmetries.

\vskip2pt
\noindent
{\bf Effective field theory:} Let us now reproduce this physics from an effective field theory. We can again trivialize conservation of the magnetic current by writing
\be
K_{\mu\nu} = \partial_{(\mu}\partial_{\nu)_T}\phi\,.
\ee
We now want to find an action for $\phi$ from which the (double) conservation of $J_{\mu\nu}$ follows. The appearance of the gauge-invariant field strength itself in the anomaly equation~\eqref{eq:confscalaranom} suggests that the relevant EFT will be Maxwell-like, and indeed this is the case. We consider the EFT
\be
S = -\frac{1}{2}\int\rd^D x \left(K_{\mu\nu}K^{\mu\nu}+\frac{1}{\Lambda^{D}}K_{\mu\nu}^4+\cdots\right)\,,
\label{eq:confscalarEFT}
\ee
where $K_{\mu\nu}^4$ stands schematically for all  index contractions, and $\Lambda$ is an energy scale (not to be confused with the gauge parameter).
Notice that the kinetic term for $\phi$ that comes from \eqref{eq:confscalarEFT} is higher derivative 
\be
S = -\int\rd^D x \left(\frac{D-1}{2D}\phi\,\square^2\phi+\cdots\right)\, .
\label{eq:confscalarfreeplusdots}
\ee
We can then understand the presence of the electric current $J_{\mu\nu}$ as a consequence of the fact $K_{\mu\nu}$, can be coupled a background gauge field by promoting it to
\be 
K_{\mu\nu} \to {\cal K}_{\mu\nu}  \equiv \partial_{(\mu} \partial_{\nu)_T} \phi -A_{\mu\nu}\,,
\label{eq:topologicalmagCS}
\ee
which is then invariant under the gauge transformations
\begin{align}
\delta\phi &= \xi\, , \\
\delta A_{\mu\nu} &= \partial_{(\mu} \partial_{\nu)_T}\xi\,.
\end{align}
with a scalar gauge parameter $\xi(x)$. The gauge field $A_{\mu\nu}$ will then couple to the electric current ${\cal J}_{\mu\nu}$:
\be
\begin{aligned}
{\cal J}_{\mu\nu} &\equiv \frac{\delta S}{\delta A^{\mu\nu}} \\
&= \partial_{(\mu }\partial_{\nu)_T}\phi - A_{\mu\nu} +\cdots\,,
\end{aligned}
\label{eq:confscalJdef}
\ee
which will be doubly conserved (because of the gauge transformation of $A$). The presence of this conserved current also guarantees that the action will have certain global symmetries, which we detail in the following inset.

\vspace{-4pt}
\begin{oframed}
{\small
\noindent
{\bf\normalsize Symmetries:} We can use the current $J_{\mu\nu}$ to construct ordinary Noether currents that correspond to the global symmetries of the EFT~\eqref{eq:confscalarEFT}. Let us illustrate this in the special case of the free theory. In this case, the doubly conserved electric current is
\be 
J_{\mu\nu}=\partial_{(\mu}\partial_{\nu)_T}\phi\, ,\label{eq:Jmunufootnote1}
\ee
which is doubly conserved as a consequence of the higher-derivative equations of motion. Note that, as expected, the current $J_{\mu\nu}$ in a generic EFT agrees with $K_{\mu\nu}$ in the deep IR. 

To obtain an ordinary conserved current from $J_{\mu\nu}$, we can start from \eqref{eq:Jmunufootnote1} and contract it with a conformal Killing scalar $\chi$, which satisfies $\partial_{(\mu}\partial_{\nu)_T}\chi=0$. The current
\be 
J_\mu[\chi]=J_{\mu\nu}\partial^\nu\chi+\partial^\nu J_{\mu\nu} \chi\, ,
\ee
will then be conserved.
We can then consider the free conformal scalar Lagrangian, and write it in the form
\be {\cal L}= -\frac{1}{2}J_{\mu\nu}^2=-\frac{1}{2} \left[ \partial_{(\mu}\partial_{\nu)_T}\phi\right]^2.
\ee
In this form, it is manifestly invariant under a shift by a conformal Killing scalar,
\be \delta \phi =\chi.
\ee
The Noether current for this symmetry is
\be {\cal J}^\mu=- \delta \phi \partial_\nu {\partial {\cal L}\over\partial( \partial_\mu\partial_\nu \phi)} +{\partial {\cal L}\over \partial(\partial_\mu\partial_\nu \phi) }\partial_\nu\delta\phi= -\partial^\mu\partial_\nu\phi\partial^\nu\chi+{1\over D}\partial^\mu\chi\square\phi+{(D-1)\over D}\chi \partial^\mu\square\phi  ,\ee
which gives exactly $-J^\mu[\chi]$.

The independent  conformal Killing scalars are
\be c,\ \ \ b_\mu x^\mu,\ \ \ x^2,\ee
such that these include the usual constant shift, galileon symmetry, as well as shifty by $x^2$, which is not a symmetry of the $(\partial\phi)^2$ kinetic term, but which {\it is} a symmetry of~\eqref{eq:confscalarEFT}. 

}
\end{oframed}

\vspace{-8pt}
The electric current defined by~\eqref{eq:confscalJdef} is guaranteed to be double conserved, but the (topological) magnetic current~\eqref{eq:topologicalmagCS} will no longer satisfy the conservation equation $\rd_{(K)}K = 0$, but will instead produce the anomaly~\eqref{eq:confscalaranom}. As expected, the EFT~\eqref{eq:confscalarfreeplusdots} that reproduces the general physics of this phase is manifestly non-unitary, having fourth-order equations of motion. 

\vskip4pt
\noindent
{\bf Electric-magnetic duality:} The linearized version of this effective field theory displays a version of electric-magnetic duality in $D=2$. In generic dimension,
we can think of the theory as being defined by the equations
\begin{align}
\label{eq:emscaleom}
\partial^\mu\partial^\nu J_{\mu\nu} &= 0\,,\\
{\cal Y}_{[2,1]}^T\,\partial_{\alpha}J_{\mu\nu} &= 0\,,
\label{eq:emscalbianchi}
\end{align}
which are valid at separated points, and
where $J_{\mu\nu} = K_{\mu\nu}$ in the linear theory. Recalling that $J_{\mu\nu}$ is part of the complex~\eqref{eq:CSsKcomp}, equation~\eqref{eq:emscalbianchi} implies that we can write\footnote{Note that, from the CFT point of view, this current is not a primary operator, so it does not appear among the multiply-conserved currents of the $\square^2$ theory classified in \cite{Brust:2016gjy}.}
\be
J_{\mu\nu} = \partial_{(\mu}\partial_{\nu)_T}\phi\,.
\ee
The equation of motion $\square^2\phi = 0$ then follows from~\eqref{eq:emscaleom}.

\noindent
{\it Two dimensions:} Something special happens in $D=2$. Tensors with the symmetry type \raisebox{2pt}{{\Yboxdim{6pt} \gyoung(~;~,~)}} vanish identically, such that the equation~\eqref{eq:emscalbianchi} is trivially satisfied in $D=2$. We must therefore additionally impose the two-derivative equation
\be
\partial_{[\alpha}\partial^\nu {\cal J}_{\mu]\nu} = - \partial_{[\alpha}\partial^\nu A_{\mu]\nu}\,,
\label{eq:twodbianchi}
\ee
in order to fix the overall normalization of the two-point function. 
The standard de Rham complex can be used to infer from~\eqref{eq:twodbianchi} that
\be
\partial^\nu J_{\mu\nu} = \partial_\mu f\,,
\ee
while the identically satisfied equation~\eqref{eq:emscalbianchi} implies that $f = \square\phi$. Then, we see that $\square^2\phi = 0$ from~\eqref{eq:emscaleom}. The takeaway is that the two equations
\begin{align}
\label{eq:Anom2DScalar0}
\partial_\mu\partial_\nu J^{\mu\nu} &= 0\,,\\
\partial_\mu \partial_{\nu} (*J)^{\mu\nu} &= 0 \implies  \partial_{[\alpha}\partial^\nu J_{\mu]\nu} = 0
\,,\label{eq:Anom2DScalar}
\end{align}
are equivalent at separated points in $D=2$, and 
completely fix this theory (along with the particular anomaly~\eqref{eq:twodbianchi} at coincident points).
Here the dualization is defined with respect to the $\mu$ index. In two dimensions, the equation~\eqref{eq:Anom2DScalar0} is equivalent to its dual, so this is a manifestly electric-magnetic duality invariant formulation of the theory. 

{Note, of course, that the choice of which equation to make anomalous at coincident points breaks electric-magnetic duality, which will appear in the contact terms of the theory.}  The equation~\eqref{eq:Anom2DScalar} is actually only true at separated points. Inside correlation functions there is an anomaly, which--for example---fixes the normalization of the two point function. The two-point function~\eqref{eq:TwoPFConfScalar} is not correct in $D=2$. Instead we find
\be
\left.\braket{J_{\mu\nu}K_{\rho\sigma}}\right|_{D=2} = \frac{2}{p^4}\left[\,p_\mu p_\nu p_\rho p_\sigma -\frac{p^2}{2} \left(p_\mu p_{(\rho}\eta_{\sigma)\nu}+ \eta_{\mu(\rho}p_{\sigma)}p_\nu \right)\right]\, ,\label{eq:TwoPF2d}
\ee
where the overall normalization is fixed by the anomaly in the equation~\eqref{eq:Anom2DScalar}, and again this correlator implies the presence of a massless scalar.

\newpage
\section{Spin-2 examples}
\label{sec:gravsec}

In the preceding section, we saw several examples of the general construction described in Section~\ref{sec:general}, and saw how the same scalar degree of freedom can realize different patterns of symmetries and anomalies, some of which cannot arise in a unitary theory. Similar constructions can lead to massless spin-$1$~\cite{Gaiotto:2014kfa,Lake:2018dqm,Cordova:2018cvg,Hofman:2018lfz} excitations, or higher $p$-form excitations.
In this section we wish to present some other examples where the gapless degree of freedom has spin two. We first review the symmetries of the EFT of a linearized graviton, before describing the higher-form symmetries of linearized conformal gravity. Conformal gravity is well-known to be non-unitary, and we will see that this is a property shared by any theory with the same generalized symmetries as (linear) conformal gravity.

\subsection{Einstein gravity}
\label{sec:GR}

We first review the anomaly structure of (linearized) Einstein gravity~\cite{Benedetti:2021lxj,Hinterbichler:2022agn,Benedetti:2023ipt} and show how it fits into the general framework described in Section~\ref{sec:general}. 

\vskip2pt
\noindent
{\bf Currents and anomalies:} The magnetic current of interest has four indices $K_{\mu_1\mu_{2}\nu_1\nu_2}$, with the symmetry type of the Weyl tensor
\be
K_{\mu_1\mu_{2}\nu_1\nu_2}
~\in~
\raisebox{5pt}{
\Yboxdim{12pt} 
\gyoung(~;~,~;~)  
}~.
\label{eq:magneticcurrentgrav}
\ee
This current is an element of the complex
\be
\Yboxdim{12pt} 
\gyoung(~) 
\xrightarrow{\,\rd_{(\xi)}\,} \,
\gyoung(~;~) 
\xrightarrow{\,\rd_{(h)}\,}
\raisebox{5pt}{
\gyoung(~;~,~;~)  
}
\xrightarrow{\,\rd_{(K)}\,}
\raisebox{10.5pt}{
\gyoung(~;~,~;~,~)
}
\xrightarrow{\,\hphantom{\rd_{(4)}}\,}
\cdots\,,
\label{eq:hcomplex}
\ee
where the various maps are defined by 
\begin{align}
\left(\rd_{(\xi)}\xi\right)_{\mu\nu} &= 2\partial_{(\mu}\xi_{\nu)_T}\,,\\
\left(\rd_{(h)}h\right)_{\mu\nu\alpha\beta} &= -3{\cal Y}_{[2,2]}^T\,\partial_\mu\partial_\alpha h_{\nu\beta}= -\partial_\mu\partial_\alpha h_{\nu\beta}+\cdots\,, \label{44eqe} \\
\left(\rd_{(K)}K\right)_{\rho\mu\nu\alpha\beta} &=3 {\cal Y}_{[2,2,1]}^T \partial_{\rho}K_{\mu\nu\alpha\beta}= \partial_{\rho}K_{\mu\nu\alpha\beta}+\cdots\,\ .
\end{align}
Using this complex, the conservation condition that is satisfied by the magnetic current is
 \be \rd_{(K)}K = 0\, ,\ee
which is the same form as the traceless part of the Bianchi identity for the Weyl tensor.

The gauge field that couples to the electric current is $A_{\mu\nu\alpha\beta}$ and also has the symmetries of the Weyl tensor.  It is part of the complex
\be
\Yboxdim{12pt} 
\raisebox{5pt}{
\gyoung(~,~) 
}
\xrightarrow{\phantom{\,\rd_{(\Lambda)}\,}}
\raisebox{5pt}{
\gyoung(~;~,~) 
}
\xrightarrow{\,\rd_{(\Lambda)}\,} \,
\raisebox{5pt}{
\gyoung(~;~,~;~)  
}
\xrightarrow{\,\rd_{(A)}\,}
\raisebox{11.5pt}{
\gyoung(~;~,~;~,~;~)  
}
\xrightarrow{\,\rd_{(F)}\,}
\cdots\,.
\label{eq:gravcomplex}
\ee
Here the gauge parameter is a mixed-symmetry tensor $\Lambda_{\mu\nu\alpha}$, 
and the differential operators are defined explicitly as
\begin{align}
\left(\rd_{(\Lambda)}\Lambda\right)_{\mu\nu\alpha\beta} &= 12{\cal Y}_{[2,2]}^T\partial_{\mu}\Lambda_{\alpha\beta\nu} = \partial_{\mu}\Lambda_{\alpha\beta\nu}+\cdots\,,\\
\left(\rd_{(A)}A\right)_{\mu\nu\rho\alpha\beta\sigma} &= 18{\cal Y}_{[2,2,2]}^T\partial_\rho\partial_\sigma A_{\mu\nu\alpha\beta} = \partial_\rho\partial_\sigma A_{\mu\nu\alpha\beta}+\cdots\,,
\label{eq:gravfieldstr}
\end{align}
so that under a gauge transformation the field transforms as\footnote{Note that the fact that $\Lambda_{\alpha\beta\nu}$ is itself the derivative of another tensor in~\eqref{eq:gravcomplex} implies that the gauge transformation is reducible, any gauge parameter that is the symmetric derivative of a $2$-form will automatically have $\delta A=0$.}
\be
\delta A_{\mu\nu\alpha\beta} =  \partial_{\mu}\Lambda_{\alpha\beta\nu}+\cdots\,,
\ee
 and the gauge-invariant field strength is, 
 \be F = \rd_{(A)}A\, .\ee
This encodes the fact that the electric current is ordinarily conserved,
\be \partial_\mu J^{\mu\nu\alpha\beta} = 0.\ee

In the presence of this background gauge field, the electric current continues to be conserved, while the magnetic current satisfies the anomaly equation
\be
\begin{aligned}
3{\cal Y}^T_{[2,2,1]}\partial_{[\rho} {\cal K}_{\mu\nu]\alpha\beta} = -{\cal C}_{\mu\nu\rho\alpha\beta} \,,
\end{aligned}
\label{eq:dualcons3}
\ee
where the anomaly can be written in terms of the gauge-invariant field strength of $A$ as\footnote{The slightly strange normalizations are chosen for convenience when matching to the EFT description. Note that the apparently singular behavior in $D=5$ is due to the fact that the field strength~\eqref{eq:gravfieldstr} vanishes identically in five dimensions, and so it is not possible to write the anomaly in terms of it. Nevertheless, it is possible to write it in terms of~\eqref{eq:Ctensorgrav} by dividing by $D-5$.}
\be
{\cal C}_{\mu\nu\rho\alpha\beta} =- \frac{(D-3)}{8(D-5)}\,\partial^\sigma F_{\mu\nu\rho\alpha\beta\sigma} =3{\cal Y}^T_{[2,2,1]}\partial_{\rho} C_{\mu\nu\alpha\beta}\,,
\ee
which also can be written in terms of the auxiliary tensor
\be
C_{\mu\nu\alpha\beta} =\frac{D-3}{D-4}{\cal Y}_{[2,2]}^T\left(
\partial^\rho \partial_\mu A_{\nu\alpha\beta \rho}- \frac{1}{4} \square A_{\mu\nu\alpha\beta}\right)\,.
\label{eq:Ctensorgrav}
\ee
As in the scalar examples, the symmetries/anomalies of this system completely fix the two point function involving the electric and magnetic currents. Explicitly, it takes the form~\cite{Hinterbichler:2022agn}.
\begin{align}
\label{eq:current2ptgenD}
\braket{J_{\mu_1\mu_2\nu_1\nu_2}K_{\alpha_1\alpha_2\beta_1\beta_2}}=  {\cal P}\,&\bigg[
 \frac{9(D-4)}{D-3}\frac{p_{\mu_1}p_{\mu_2}p_{\alpha_1}p_{\alpha_2}\eta_{\nu_1\beta_1}\eta_{\nu_2\beta_2}}{p^2}
 \\&~~
+\frac{3}{4}p^2\eta_{\mu_1\alpha_1}\eta_{\mu_2\alpha_2}\eta_{\nu_1\beta_1}\eta_{\nu_2\beta_2} - 3 p_{\mu_1}p_{\alpha_1}\eta_{\mu_2\alpha_2}\eta_{\nu_1\beta_1}\eta_{\nu_2\beta_2}\bigg]\,,\nonumber
\end{align}
where ${\cal P} \equiv {\cal Y}_{[2,2]}^T {\cal Y}_{[2,2]}^T $ is a Young projector onto the tableau that has the symmetries of a 
symmetric product of Weyl tensors
{\small
\be
 \raisebox{1.5ex}{\Yboxdimx{13.5pt}
\Yboxdimy{13.5pt}\gyoung({{\hspace{.1em}\mu_1}};{{\hspace{.1em}\nu_1}},{{\hspace{.09em}\mu_2}};{{\hspace{.09em}\nu_2}})}~\medotimes\, \raisebox{1.5ex}{\Yboxdimx{13.5pt}
\Yboxdimy{13.5pt}\gyoung({{\hspace{.075em}\alpha_1}};{{\hspace{.05em}\beta_1}},{{\hspace{.075em}\alpha_2}};{{\hspace{.05em}\beta_2}})}\, .
\label{eq:weylweylprojectortraceless}
\ee
\vskip.5pt
}
\noindent
Note that only the first term of~\eqref{eq:current2ptgenD} is nonlocal. 
Performing a spectral decomposition on the two-point function~\eqref{eq:current2ptgenD} implies that there is a gapless spin-2 mode---the linearized graviton---in the spectrum~\cite{Hinterbichler:2022agn}. From this perspective, the graviton is a Goldstone for a (higher-form) 
symmetry, which is nonlinearly realized.

\vskip2pt
\noindent
{\bf Effective field theory:} We now want to review how this structure of anomalies is reproduced in effective field theory~\cite{Hinterbichler:2022agn,Benedetti:2023ipt}. As before, the strategy is to trivialize the magnetic conservation condition by writing the magnetic current in terms of a potential field
\be
K_{\mu_1\mu_1\nu_1\nu_2} = -3{\cal Y}_{[2,2]}^T\partial_{\mu_1}\partial_{\nu_1}h_{\mu_2\nu_2}\,,\label{mehgrrive}
\ee
where $h_{\mu\nu}$ is a traceless symmetric tensor.
Of course this is just the linearized Weyl tensor where $h_{\mu\nu}$ is the linearized metric in a gauge where $h_{\ \mu}^\mu = 0$.
Notice that the complex~\eqref{eq:hcomplex} implies that this tensor is invariant under the gauge transformation
\be
\delta h_{\mu\nu} = 2\partial_{(\mu}\xi_{\nu)_T}\,,
\ee
which is nothing but the linearized diffeomorphism invariance that preserves $h_{\ \mu}^\mu = 0$.
 We see that gauge invariance in this language is an inevitable consequence of choosing to write one of the currents locally in terms of field variables that trivialize its conservation.\footnote{We would have seen a similar phenomenon in the scalar examples if we had decided to trivialize the electric current conservation, which would have introduced a $(D-2)$-form field carrying the gapless mode.}

We now want to construct an action whose equation of motion implies the conservation of the electric current $J_{\mu_1\mu_1\nu_1\nu_2}$. This current can be seen as the Noether current for the shift
\be
\delta h_{\mu\nu} = 2\partial^\alpha\Lambda_{\alpha(\mu\nu)}\,,
\label{eq:lhfsymm}
\ee
where $\Lambda$ is a traceless mixed-symmetry tensor and 
we must require that $K[\delta h]_{\mu\nu\alpha\beta} =0$ in order for this to be a symmetry. (This condition is the analogue of the $1$-form symmetry in electromagnetism requiring that we shift by a flat connection.) 

As in the galileon superfluid example discussed in Section~\ref{sec:galSF}, the most naive action built directly from $K$ will have fourth-order equations of motion, and the wrong anomaly. It is, however, possible to construct a kinetic term with lower-order equations of motion: 
\be
S = -  \int\rd^D x \,\frac{1}{2}h^{\mu\nu}\Big(
\square h_{\mu\nu}-2\partial^\alpha\partial_{(\mu} h_{\nu)\alpha}
\Big)\,,
\label{eq:tracelesslinearizedEH}
\ee
whose equation of motion sets the traceless linearized Einstein tensor to zero $G_{(\mu\nu)_T} = 0$. Note that this is a non-standard presentation of linearized gravity, written in terms of an identically traceless tensor (see section 2.4 of \cite{Bonifacio:2015rea}).  The electric current in this linearized theory is also the Weyl tensor
\be
W_{\mu_1\mu_1\nu_1\nu_2} = -3{\cal Y}_{[2,2]}^T\partial_{\mu_1}\partial_{\nu_1}h_{\mu_2\nu_2}\,,
\label{eq:linearizedweyl}
\ee
which is conserved on-shell. As before, the electric current happens to coincide with $K$, but this is again an accident of the linearized theory. We can introduce interactions, which are naturally built from powers of $K$. The electric current will then receive corrections (but will continue to be conserved on-shell), while the magnetic current will be unchanged.  These will be interactions of the ``pseudo-linear'' type~\cite{Wald:1986bj,Li:2015vwa,Chatzistavrakidis:2016dnj,Bai:2017dwf,Bonifacio:2018van}. It is an open question how to extend this framework to describe the non-abelian interactions of full Einstein--Hilbert gravity.

We can gauge the symmetry~\eqref{eq:lhfsymm} by introducing  $A_{\mu\nu\alpha\beta}$ into the action~\eqref{eq:tracelesslinearizedEH} as
\be
\begin{aligned}
S = \int\rd^Dx \bigg(&\frac{1}{2}\partial_\alpha h_{\mu\nu}\partial^\alpha h^{\mu\nu}-\partial^\mu h_{\mu\alpha}\partial_\nu h^{\nu\alpha}
-A_{\mu\nu\rho\sigma}\partial^\mu\partial^\rho h^{\nu\sigma}
\\&\hspace{1.75cm}-\frac{D-3}{8(D-4)} \partial^\mu A_{\nu\rho\sigma\alpha}\partial_\mu A^{\nu\rho\sigma\alpha}+\frac{D-3}{2(D-4)}\partial^\mu A_{\nu\rho\sigma\mu} \partial_\alpha A^{\nu\rho\sigma\alpha}\bigg)\,,
\end{aligned}
\label{eq:genDaction}
\ee
which is invariant under the following gauge transformations
\begin{align}
\delta_\Lambda A_{\mu\nu\alpha\beta} & = 12{\cal Y}_{[2,2]}^T\partial_{\mu}\Lambda_{\alpha\beta\nu} = \partial_\mu \Lambda_{\nu\beta\alpha}+\cdots\,\\
\delta_\Lambda h_{\mu\nu} &= \frac{6(D-3)}{D-2}\partial^\alpha\Lambda_{\alpha(\mu\nu)}\,.
\end{align}
This action can be written more simply as
\be 
S = -\frac{1}{2}\int \rd^dx \bigg(h^{\mu\nu}\mathcal{G}_{\mu\nu} + A^{\mu\nu\rho\sigma}\mathcal{J}_{\mu\nu\rho\sigma}\bigg)\, ,\label{eq:NiceActGeom}
\ee
where we have introduced the gauge-improved (traceless) Einstein tensor and electric current
\begin{align}
\label{eq:gaugeinveins}
{\cal G}_{\mu\nu} &= \square h_{\mu\nu} - 2\partial^\alpha\partial_{(\mu}h_{\nu)_T\alpha}
-\partial^\alpha\partial^\beta A_{\mu\alpha\nu\beta}\,,\\
{\cal J}_{\mu\nu\alpha\beta} &= \frac{1}{4}W_{\mu\nu\alpha\beta} -C_{\mu\nu\alpha\beta}\,.
\label{eq:Jingend}
\end{align}
In the absence of the gauge field, the electric current is just the linearized Weyl tensor, and in the gauge-invariant version
 the $C$ tensor is the same as in~\eqref{eq:Ctensorgrav}.
In the linearized theory, the magnetic current is equal to the electric current, so we also have
\be
{\cal K}_{\mu\nu\alpha\beta} = \frac{1}{4}W_{\mu\nu\alpha\beta} -C_{\mu\nu\alpha\beta}\,.
\label{eq:Kingend}
\ee
One can then check that the antisymmetric derivative of this current reproduces the anomaly~\eqref{eq:dualcons3}, where we have
\be
{\cal C}_{\mu\nu\rho\alpha\beta} =  3{\cal Y}^T_{[2,2,1]}\partial_{\rho} C_{\mu\nu\alpha\beta}\,.
\label{eq:dimeqsgravity}
\ee

Gauge-invariant interactions of the pseudo-linear type are introduced by including higher-order contractions of ${\cal K}$. Then, in the full theory ${\cal K}$ (and hence its anomaly) will remain unchanged, while the electric current ${\cal J}_{\mu\nu\alpha\beta}$ will be modified by the interactions.

\subsection{Conformal gravity}
\label{sec:confgrav}
We now turn to (linearized) conformal gravity. This theory shares many features with the conformal scalar example encountered in~\eqref{sec:confscal}: both its magnetic current and electric gauge field will be governed by the same differential complex, and the mixed anomaly between conservation of the two symmetries will imply that the theory is non-unitary. Of course the non-unitarity of conformal gravity is well known, but this perspective casts this fact as an inevitable consequence of symmetry, and indeed exposes the (higher-form) symmetries of conformal gravity itself. As a byproduct of our analysis, we elucidate the fate and structure of electric-magnetic duality for linearized conformal gravity in four spacetime dimensions. (See Appendix~\ref{app:confgrav} for more details about  (linearized) conformal gravity.)

\vskip2pt
\noindent
{\bf Currents and anomalies:} We start by defining the magnetic current in the same way as in linearized Einstein gravity~\eqref{eq:magneticcurrentgrav}. The index symmetries of $K_{\mu_1\mu_2\nu_1\nu_2}$ are identical and it fits into the same differential complex~\eqref{eq:hcomplex}.
The essential difference with the linearized GR case is that we take the gauge field $A_{\mu\nu\alpha\beta}$ to be part of this same differential complex,
\be
\Yboxdim{12pt} 
\gyoung(~) 
\xrightarrow{\,\hphantom{\rd_{(\xi)}}\,} \,
\gyoung(~;~) 
\xrightarrow{\,\rd_{(\Lambda)}\,}
\raisebox{5pt}{
\gyoung(~;~,~;~)  
}
\xrightarrow{\,\rd_{(A)}\,}
\raisebox{10.5pt}{
\gyoung(~;~,~;~,~)
}
\xrightarrow{\,\rd_{(F)}\,}
\cdots\,.
\label{eq:weylcomplex}
\ee
Here the relevant maps involved can be written explicitly as
\begin{align}
\left(\rd_{(\Lambda)}\Lambda\right)_{\mu\nu\alpha\beta} &= 3{\cal Y}_{[2,2]}^T\,\partial_\mu\partial_\alpha \Lambda_{\nu\beta}\,,\\
\left(\rd_{(A)}A\right)_{\rho\mu\nu\alpha\beta} &= 6{\cal Y}_{[2,2,1]}^T\,\partial_{\rho}A_{\mu\nu\alpha\beta}\,.
\end{align}
In particular this implies that
under a gauge transformation the gauge field transforms as 
\be \delta A_{\mu\nu\alpha\beta} = {\cal Y}_{[2,2]}^T\,\partial_\mu\partial_\alpha \Lambda_{\nu\beta},\ee
and there is a gauge-invariant field strength 
\be F_{\rho\mu\nu\alpha\beta} = {\cal Y}_{[2,2,1]}^T\,\partial_{\rho}A_{\mu\nu\alpha\beta}\, .\label{Fconfgffielde}\ee
The fact that $A$ has a gauge transformation involving two derivatives implies that the electric current that it couples to satisfies the double conservation condition
\be
\partial_{\mu_1}\partial_{\nu_1}J^{\mu_1\mu_2\nu_1\nu_2} = 0 \,.
\label{eq:congrav1}
\ee

The presence of the background gauge field $A$ that sources the electric current $J$ preserves this conservation condition, but causes $K$ to satisfy the anomalous conservation equation
\be
{\cal Y}_{[2,2,1]}^T\partial_{\rho}{\cal K}_{\mu_1\mu_2\nu_1\nu_2} = - F_{\rho\mu_1\mu_2\nu_1\nu_2 }\,,
\label{eq:cganom}
\ee
where $F$ is the gauge-invariant field strength \eqref{Fconfgffielde} built out of $A$. Notice that in this case the field strength has the right index symmetries to appear on the right hand side of the anomaly equation without needing to take derivatives. In this sense, conformal gravity is  conceptually quite similar to the conformal scalar example discussed in Section~\ref{sec:confscal}.

\vskip2pt
\noindent
{\bf Two-point function and spectrum:} As in the other examples, the combined requirements of conservation of the electric current at all points, and the 't Hooft anomaly~\eqref{eq:cganom} entirely fixes the two-point functions between these currents to be 
 \be
 \begin{aligned}
\braket{J_{\mu_1\mu_2\mu_3\mu_4}K_{\nu_1\nu_2\nu_3\nu_4}} =-{\cal P}\bigg[
 &\frac{p_{\mu_1}p_{\alpha_1}\eta_{\nu_1\beta_1}}{p^4}\left(-6p_{\mu_2}p_{\beta_2}\eta_{\nu_2\alpha_2} +p^2 \eta_{\mu_2\alpha_2}\eta_{\nu_2\beta_2}\right)
 \\
 &+\frac{3}{4}\eta_{\mu_1\alpha_1}\eta_{\mu_2\alpha_2}\eta_{\nu_1\beta_1}\eta_{\nu_2\beta_2} - \frac{3 p_{\mu_1}p_{\alpha_1}\eta_{\mu_2\alpha_2}\eta_{\nu_1\beta_1}\eta_{\nu_2\beta_2}}{p^2}\\
 &+ \frac{9(D-4)}{D-3}\frac{p_{\mu_1}p_{\mu_2}p_{\alpha_1}p_{\alpha_2}\eta_{\nu_1\beta_1}\eta_{\nu_2\beta_2}}{p^4}
\bigg]\, ,
\end{aligned}
\ee
where ${\cal P}$ is the same projector as in~\eqref{eq:current2ptgenD}.
Notice that this two point function has a $p^{-4}$ pole. This pole cannot be reproduced by any unitary representation in the \KL decomposition of this correlator. We therefore conclude that any theory with this pattern of symmetries must necessarily be non-unitary. In the next section we will see that the EFT description of this physics is linearized conformal gravity, which is entirely consistent with the theory being non-unitary.

\paragraph{Effective field theory:}
We now want to see how we can reproduce the physics of this phase using effective field theory.
Once again the magnetic current's conservation can be trivialized by writing the current as~\eqref{mehgrrive}. We further notice that this current is again invariant under the shift symmetry~\eqref{eq:lhfsymm}. In this case, we must use a Maxwell-type action to realize the relevant symmetries.  This Maxwell-type action is quadratic in the magnetic current $K$,
\be 
S =\frac{1}{4}\int \rd^Dx \,K_{\mu\nu\alpha\beta}K^{\mu\nu\alpha\beta}\,.
\label{eq:linearizedweyl2}
\ee
The action~\eqref{eq:linearizedweyl2} is of course the action of linearized conformal gravity in a gauge in which the linearized Weyl symmetry is used to make $h_{\mu\nu}$ traceless, with $K$ the linearized Weyl tensor. As in the Einstein gravity case, we could include higher-order interaction terms involving powers of the Weyl tensor. This would leave the magnetic current unchanged, but would modify the electric current.

In order to understand the structure of anomalies, we want to gauge the symmetry~\eqref{eq:lhfsymm} by introducing a gauge field that couples to the electric current. We can construct a gauge-invariant action as (now writing ${\cal K}$ in terms of the Weyl tensor),
\be 
S =\frac{1}{4}\int \rd^Dx\, \big(W_{\mu\nu\alpha\beta}-A_{\mu\nu\alpha\beta}\big)\big(W^{\mu\nu\alpha\beta}-A^{\mu\nu\alpha\beta}\big)\, ,
\ee
which is just the square of the gauge-invariant magnetic current,
\be
{\cal K}_{\mu\nu\rho\sigma}= W_{\mu\nu\rho\sigma} - A_{\mu\nu\rho\sigma}\,.
\ee
This is invariant under the following gauge transformations of the fields,
\begin{align}
\delta h_{\mu\nu} &= \Lambda_{\mu\nu}\, ,\\
\delta A_{\mu\nu\alpha\beta} &= -3\mathcal{Y}_{[2,2]}^T\,\partial_\nu \partial_\beta \Lambda_{\mu\alpha}\, ,
\end{align}
where $\Lambda_{\mu\nu}$ is a symmetric traceless gauge parameter.  
In the presence of $A_{\mu\nu\alpha\beta}$, the electric current that couples to $A$ will be conserved, but the magnetic current 
now obeys the anomalous equation~\eqref{eq:cganom}. So we see that this effective field theory precisely reproduces the structure of symmetries and anomalies that we want. If desired, we can incorporate interactions by adding higher powers of ${\cal K}$ to the action.

As anticipated, this effective field theory involves a higher-derivative action, and it is known to be non-unitary, consistent with our statement that the symmetries~\eqref{eq:congrav1} and~\eqref{eq:cganom} cannot arise in a unitary EFT.

\subsubsection{Charges}

Here we discuss some conserved currents and charges that arise from the conserved electric current in conformal gravity. 
$W_{\mu\nu\alpha\beta}$ is a double conserved ($\partial^\nu\partial^\beta W_{\mu\nu\alpha\beta}=0$) traceless operator, with the symmetries of the Young diagram \raisebox{3pt}{\Yboxdim{6pt}
\gyoung(~;~,~;~)}.
From this we can construct 2-form conserved currents $J_{\mu\nu}[\xi]$ by contracting with a conformal killing vector $\xi^\mu$ satisfying the conformal Killing equation $\partial_{(\mu}\xi_{\mu)_T}=0$  as follows,
\be 
J_{\mu\nu}[\xi]=W_{\mu\nu\alpha\beta}\partial^\alpha \xi^\beta-2\partial^\alpha W_{\mu\nu\alpha\beta}\, \xi^\beta\, .
\label{eq:Jdef}
\ee
These currents are conserved: $ \partial^\nu J_{\mu\nu}[\xi]=0$, as we show in the following inset.

\vspace{-4pt}
\begin{oframed}
{\small
\noindent
{\bf\normalsize Conservation of $\boldsymbol{J_{\mu\nu}}[\xi]$:} 
To see that $J_{\mu\nu}[\xi]$ defined in~\eqref{eq:Jdef} is conserved, act with $\partial^\nu$ and use the fact that
 $W$ is double conserved to give
\be \partial^\nu{ J}_{\mu\nu}[\xi]=\partial^\nu W_{\mu\nu\alpha\beta}\partial^\alpha \xi^\beta+W_{\mu\nu\alpha\beta}\partial^\nu\partial^\alpha \xi^\beta-2\partial^\alpha W_{\mu\nu\alpha\beta}\, \partial^\nu\xi^\beta\,.\label{Wdconsjeeqe}\ee
We can treat $\partial^\mu \xi^\nu$ as anti-symmetric---since the symmetric traceless part vanishes by the conformal Killing equation---and the trace part vanishes upon contracting with the traceless $W$.  So, the second term, which has two derivatives on $\xi$, can be written as
\be W_{\mu\nu\alpha\beta}\partial^\nu\partial^\alpha \xi^\beta= -W_{\mu\beta\alpha\nu}\partial^\alpha\partial^\nu  \xi^\beta \,, \ee
which then vanishes due to the symmetry of the two derivatives and the antisymmetry of $W$.  
Now, looking at the two terms in \eqref{Wdconsjeeqe} with one derivative on $\xi$, we can write them as
\be 
\partial^\nu W_{\mu\nu\alpha\beta}\partial^\alpha \xi^\beta -2\partial^\alpha W_{\mu\nu\alpha\beta}\, \partial^\nu\xi^\beta= \partial^\nu \left( W_{\mu\nu\alpha\beta}-2W_{\mu\alpha\nu\beta} \right) \partial^\alpha \xi^\beta\,.\ee
Again treating $\partial^\mu \xi^\nu$ as anti-symmetric, we can manipulate this as follows,
\be
\begin{aligned}
\partial^\nu \left( W_{\mu\nu\alpha\beta}-2W_{\mu\alpha\nu\beta} \right)\partial^\alpha \xi^\beta 
&= \partial^\nu \left( W_{\mu\nu\alpha\beta}-W_{\mu\alpha\nu\beta} +W_{\mu\beta\nu\alpha} \right) \partial^\alpha \xi^\beta\\
&=\partial^\nu \left( W_{\mu\nu\alpha\beta}+W_{\mu\alpha\beta\nu} +W_{\mu\beta\nu\alpha} \right) \partial^\alpha \xi^\beta \, ,
\end{aligned}
\ee
which now vanishes upon using the mixed-symmetry condition $W_{\mu[\nu\alpha\beta]}=0$.

}
\end{oframed}

Due to the fourth order nature of the conformal gravity field equations, solutions to Einstein gravity are also solutions to conformal gravity, but the converse is not true: it would be interesting to study linearized black hole solution in conformal gravity that are not also solutions of Einstein gravity~\cite{Riegert:1984zz,Mannheim:1988dj}, to see if and how they carry these charges, and study their topology, along the lines of~\cite{Hinterbichler:2015nua,Hinterbichler:2022agn,Hull:2024xgo}.

\subsubsection{Electric-magnetic duality}
In this section, we describe the electric-magnetic-like duality invariance of linearized conformal gravity in $D=4$, which is quite similar to the conformal scalar in $D=2$ discussed in Section~\ref{sec:confscal}. Recall that linearized conformal gravity 
is on-shell equivalent to the equations
\begin{align}
\label{eq:weylwaveeq}
\partial_{\mu_1}\partial_{\nu_1} J^{\mu_1\mu_2\nu_1\nu_2} & =0\,, \\
\partial^{\mu_1}(*J)_{\mu_1\cdots \mu_{D-2}\nu_1\nu_2} &=0 \implies {\cal Y}_{[2,2,1]}^T\partial_{[\mu_1}J_{\mu_1\mu_3]\nu_1\nu_2}  =0\,, 
\label{eq:weybianchi}
\end{align}
where the tensor $J_{\mu_1\mu_2\nu_1\nu_2}$ has the symmetries of the window-shaped tableau \raisebox{3pt}{\Yboxdim{6pt}
\gyoung(~;~,~;~)}.
The Hodge dual is defined with respect to the first two indices of $J$ so that
 $(*J)_{\mu_1\cdots \mu_{D-2}\nu_1\nu_2} = \tfrac{1}{2}\epsilon_{\mu_1\mu_2\cdots \mu_{D-2}\alpha\beta}J^{\alpha\beta}_{\hspace{12pt} \nu_1\nu_2}$. 
In general dimension $J_{\mu_1\mu_2\nu_1\nu_2}$ is nothing but the Weyl tensor, and we can use the fact that it fits into the complex~\eqref{eq:weylcomplex} along with~\eqref{eq:weybianchi} write
\be
 J_{\mu_1\mu_2\nu_1\nu_2} = {\cal Y}_{[2,2]}^T\,\partial_{\mu_1}\partial_{\nu_1}h_{\mu_2\nu_2}\,,
\ee 
so that~\eqref{eq:weylwaveeq} is the wave equation satisfied by the conformal graviton~\eqref{eq:confgravEOM}.

In $D=4$, the equation~\eqref{eq:weybianchi} is trivially satisfied, and so the equation~\eqref{eq:weylwaveeq} is not sufficient to completely fix the theory. In this case, we have to additionally impose
\be
\partial^{\mu_1}\partial^{\nu_1}(*J)_{\mu_1\mu_2\nu_1\nu_2} = 0\,.
\label{eq:extraeq}
\ee
The equations~\eqref{eq:weylwaveeq} and~\eqref{eq:extraeq} are sufficient to fix the two-point function of the current to be that of conformal gravity. We can also reproduce the equations of motion by noting that we can equivalently write this equation as
\be
\partial_{[\mu_1}\partial^{\nu_1}J_{\mu_1\mu_3]\nu_1\nu_2} = 0\,.
\label{eq:extraeq2}
\ee
Then, the fact that the
mixed-symmetry tensor appearing in this equation fits into the complex~\eqref{eq:CSsKcomp} implies that we can write
\be
\partial^{\nu_1}J_{\mu_1\mu_2\nu_1\nu_2} = \partial_{[\mu_1}S_{\mu_2]\nu_2}\,,
\ee
where $S$ is a symmetric tensor. We can also use the trivial equation~\eqref{eq:weybianchi} to deduce that
\be
 J_{\mu_1\mu_2\nu_1\nu_2} = {\cal Y}_{[2,2]}^T\,\partial_{\mu_1}\partial_{\nu_1}h_{\mu_2\nu_2}\,.
\ee
Combining these equations, we find that 
\be
S_{\mu\nu} = -2\square h_{\mu\nu} +2\partial^\alpha\partial_{(\mu}h_{\nu)\alpha} \,.
\ee
Then, applying~\eqref{eq:weylwaveeq}, we find the linearized wave equation of conformal gravity. All together, we see that in $D=4$, linearized conformal gravity is equivalent to the two equations
\begin{align}
\partial_{\mu_1}\partial_{\nu_1} J^{\mu_1\mu_2\nu_1\nu_2} & =0\,,\\
\partial_{\mu_1}\partial_{\nu_1}(*J)^{\mu_1\mu_2\nu_1\nu_2} &= 0\,,
\label{eq:anomaleqconfg}
\end{align}
which are clearly Hodge duals of each other. The same conclusion was reached by~\cite{Snethlage:2021lsf} via a slightly different chain of reasoning.\footnote{See~\cite{Bekaert:2002dt,Hinterbichler:2016fgl} for discussions of the duality-invariant formulation of linearized Einstein gravity.} As in the scalar case, the anomaly appearing in~\eqref{eq:anomaleqconfg} at coincident points (which we have not written explicitly) is necessary to fully fix the normalization of the two-point function.

\newpage
\section{Conclusions}
\label{sec:conclusion}

\vspace{-6pt}
We have explored the properties of relativistic gapless phases defined by a pair of generalized conserved currents with a mixed anomaly between them. We have given a unified description of these systems, and have considered several examples, both old and new, and shown how they fit into the general framework.
Perhaps the most conceptually nontrivial aspect of these examples is that some current and anomaly structures cannot possibly arise in a unitary (Lorentzian) quantum field theory: these symmetries are impossible.

Along the way, we elucidated the structure of symmetries in conformal gravity, giving a definition of the linearized theory in terms of higher-form symmetries. This theory is one with impossible symmetries, reflecting the well-known fact that conformal gravity is non-unitary. We also considered some aspects of the conserved charges in this theory, and elucidated features of its electric-magnetic duality.
A similar construction should exist for massless higher spin fields described by the Fronsdal action~\cite{Fronsdal:1978rb}, as well as the Fradkin--Tseytlin fields~\cite{Fradkin:1985am,Segal:2002gd} (which are higher-spin generalizations of linearized conformal gravity), and would give a new perspective on these theories.  In these cases, there is a generalized Weyl tensor satisfying conservation equations, whose anomalies should characterize the theory, and give an interpretation of each massless high spin as a Goldstone boson.

A number of questions are raised by this construction. Philosophically, one could ask why some symmetries are impossible, and whether we could predict a priori when currents are incompatible.  One avenue to further elucidate this structure would be to connect these statements to more familiar theorems that forbid the presence of certain conserved currents: either the Weinberg--Witten~\cite{Weinberg:1980kq,Benedetti:2022zbb} or Maldacena--Zhiboedov~\cite{Maldacena:2011jn} theorems (see also \cite{Boulanger:2013zza,Alba:2015upa} for higher-dimensional generalizations).  By invoking standard unitarity, we had in mind the relativistic setting, where the constraints of Lorentz invariance imply a certain rigidity, but for condensed matter applications, it would be useful to relax this assumption and study whether these patterns of symmetries can be realized by physical systems that are interpreted as statistical ensembles (i.e. as Euclidean field theories).  In any setting, it would be interesting to understand if there is a more abstract and general characterization of impossible symmetries.

There is a complementary perspective on impossible symmetries that comes from recent developments in the study of positivity bounds. By studying scattering amplitudes away from the forward limit, it is possible to obtain two-sided bounds on the Wilson coefficients of EFTs~\cite{Bellazzini:2020cot,Tolley:2020gtv,Caron-Huot:2020cmc}. One way of phrasing the lesson is that large hierarchies in dimensionless parameters in an EFT usually cannot descend from a standard Lorentz invariant UV completion.  As a concrete example, scalar field theories with amplitudes that are parametrically softer than $E^2$ in the forward limit cannot arise from such a completion.  In particular, this rules out models of galileons~\cite{Nicolis:2008in} arising in this way. This is, in a sense, another statement about ``impossible" symmetries, as the galileon realizes a particular low-energy symmetry which is responsible for the softness of its amplitudes. It would be very interesting if one could understand these violations of positivity by using the extended operators or higher-form currents present in the theory of a galileon, or in more general EFTs.

We are just beginning to understand the interplay between generalized symmetries and the structure of EFT. A number of novel and interesting phenomena have already been uncovered, and we expect that there is a wealth of additional insights to be gained.

\paragraph{Acknowledgements:} We would like to thank Clifford Cheung and Diego Hofman for helpful discussions.  KH acknowledges support from DOE grant DE-SC0009946.  AJ is supported in part by DOE (HEP) Award DE-SC0009924. GM is supported in part by the Simons Foundation grant 488649 (Simons Collaboration on the Nonperturbative Bootstrap) and the Swiss National Science Foundation through the project 200020 197160 and through the National Centre of Competence in Research SwissMAP. GM also acknowledges support from NSF grant PHY-1748958 for participation in a KITP workshop. Part of this work was performed in part at the Aspen Center for Physics, which is supported
by NSF grant PHY-2210452.

\appendix

\section{Conformal gravity }
\label{app:confgrav}

Here we review some aspects of both nonlinear and linearized conformal gravity that are useful in the main text.

\subsection{Nonlinear conformal gravity}

We begin by considering nonlinear Weyl-squared gravity. The action of this theory is given by\footnote{Given the definition of the Weyl tensor, this action can equivalently be written in terms of the Riemann tensor as
\be
S_{W^2} = -\frac{1}{8}\int \rd^Dx \sqrt{g
}\, \left(R_{\mu\nu\alpha\beta}R^{\mu\nu\alpha\beta}-\frac{4}{D-2}R_{\mu\nu}R^{\mu\nu}+\frac{2}{(D-1)(D-2)}R^2\right)\, .\label{eq:SconfAp2}
\ee
}
\be
S_{\rm W^2} = -\frac{1}{8}\int \rd^Dx \sqrt{g
}\, W_{\mu\nu\rho\sigma}W^{\mu\nu\rho\sigma}\, ,\label{eq:SconfAp}
\ee
where $W_{\mu\nu\rho\sigma}$ is the Weyl tensor, which is the traceless part of the Riemann tensor, 
\begin{align} 
W_{\mu\nu\rho\sigma}& =R_{\mu\nu\rho\sigma}- \frac{2}{D-2}\left(g_{\mu[\rho}R_{\sigma]\nu}-\eta_{\nu[\rho}R_{\sigma]\mu}\right)+\frac{2}{(D-1)(D-2)}\eta_{\mu[\rho}\eta_{\sigma]\nu}R\, \nn\\
&= R_{\mu\nu\rho\sigma}-2(g_{\mu[\rho}S_{\sigma]\nu}-g_{\nu[\rho}S_{\sigma]\mu})\, .
\end{align}
In the second line, we have expressed it in terms of the
Schouten tensor, which is defined as 
\be 
S_{\mu\nu} = \frac{1}{D-2}\left(R_{\mu\nu}-\frac{1}{2(D-1)}g_{\mu\nu}R\right)\, .
\ee
The Weyl tensor shares the same index symmetries as the Riemann tensor, $W_{\mu\nu\rho\sigma} = -W_{\nu\mu\rho\sigma} = -W_{\mu\nu\sigma\rho} = W_{\rho\sigma\mu\nu}\,$,
and the algebraic Bianchi identity $W_{[\mu\nu\rho]\sigma}=0$, and it is completely traceless i.e., it has the symmetries of the traceless $[2,2]$ tableau  \raisebox{4pt}{\tiny\Yboxdim{8pt}
\gyoung(\mu;\sigma,\nu;\rho)}.

The divergence of the Weyl tensor can be written as
\be 
\nabla^\mu W_{\mu\nu\rho\sigma} = -\frac{(D-3)}{(D-2)}C_{\nu\rho\sigma}\, ,\label{eq:WeylDiv}
\ee
where $C_{\nu\rho\sigma}$ is the Cotton tensor, which can be written as
\be
C_{\nu\rho\sigma} = 2(D-2) \nabla_{[\sigma}S_{\rho]\nu}
= -2 \nabla_{[\rho} R_{\sigma]\nu}- \frac{1}{D-1}g_{\nu[\rho}\nabla_{\sigma]} R \, .
\ee
The Cotton tensor is fully traceless and has the symmetry of a $[2,1]$ tableau  \raisebox{4pt}{\tiny\Yboxdim{8pt}
\gyoung(\rho;\nu,\sigma)}.

The Weyl tensor with one index raised is invariant under Weyl transformations, that is, if we define
\be
g_{\mu\nu} = \Omega^2(x) \tilde g_{\mu\nu}\,,
\ee
then the Weyl tensors of these two metrics are equivalent,
\be
W^\mu_{\ \ \nu\alpha\beta}=\tilde W^\mu_{\ \ \nu\alpha\beta}\,.
\ee
This implies that in $D=4$, the action~\eqref{eq:SconfAp} is invariant under Weyl transformations $\delta g_{\mu\nu} \mapsto \Omega^2(x)g_{\mu\nu}$ because the transformation of the inverse metrics needed to contract indices cancel against the measure. 
Taking $\Omega(x) = e^{\sigma(x)}$, infinitesimally the Weyl transformation reads 
\begin{align}
\delta g_{\mu\nu} &= 2\sigma(x)g_{\mu\nu}\, . \label{infweyle}
\end{align}
Due to this Weyl invariance, In $D=4$, we call \eqref{eq:SconfAp} conformal gravity. 
In other dimensions, the action \eqref{eq:SconfAp} is {\it not} Weyl invariant.\footnote{
In higher dimensions, there are multiple independent Weyl invariants~\cite{Boulanger:2018rxo}, so it is not immediately obvious what one should mean by conformal gravity for $D>4$.  For $D$ even, there is a natural candidate; it is a $D$-derivative theory that when expanded around (A)dS contains a massless, partially massless, and ${D-4\over 2}$ massive, spin-2 degrees of freedom (see e.g.,~\cite{Joung:2019wwf}).  Its linearized action is the linearization of $\sim\int \rd^Dx\, W_{\mu\nu\alpha\beta}\square^{D-4\over 2} W^{\mu\nu\alpha\beta} $, which is a spin 2 Fradkin--Tseytlin field~\cite{Fradkin:1985am,Segal:2002gd}. } 

The equation of motion derived from the action~\eqref{eq:SconfAp} in $D=4$ is $B_{\mu\nu} = 0$, where $B$ is the Bach tensor\footnote{Note that in the literature, there exist different definitions for the Bach tensor in general dimension $D>4$. One choice is to define the Bach tensor as the tensor that is set to zero by the equations of motion derived from the action~\eqref{eq:SconfAp},
\begin{align}
\bar{B}_{\mu\nu} \equiv -2{\delta S_{W^2}\over \delta g^{\mu\nu}} = -\nabla^{(\rho}\nabla^{\sigma)}W_{\mu\rho\nu\sigma}-{1\over 2}W\indices{_\mu^{\rho\sigma\alpha}}W_{\nu\rho\sigma\alpha}-\frac{1}{D-2}R^{\rho\sigma}W_{\mu\rho\nu\sigma}+\frac{1}{8}g_{\mu\nu}W_{\mu\nu\rho\sigma}^2\, .
\end{align}
This tensor is divergenceless ($\nabla^\mu \bar{B}_{\mu\nu}=0$) in any dimension, as required by diffeomorphism invariance, but only traceless in $D=4$.

Another commonly used definition of the Bach tensor in general dimension is $B_{\mu\nu}$ that we presented in the main text in equation \eqref{eq:BachAp1} and \eqref{eq:BachAp2}.
The tensor $B_{\mu\nu}$ in \eqref{eq:BachAp2} is now traceless in any dimension, but its divergence is
\be 
\nabla^\mu B_{\mu\nu} = (D-4)S^{\alpha\beta}\left(\nabla_\nu S_{\alpha\beta}-\nabla_\alpha S_{\nu\beta}\right)\, ,
\ee
which vanishes in $D=4$.

These two definitions coincide in $D=4$ (in showing this, it is useful to add the variation of the Gauss--Bonnet term, which vanishes in $D=4$). Moreover, in $D=4$ the Bach tensor is Weyl covariant,
\be 
\delta B_{\mu\nu} = 2\sigma B_{\mu\nu}\, ,
\ee
under infinitesimal Weyl transformations $\delta g_{\mu\nu}=2\sigma g_{\mu\nu}$.}
\bea
B_{\mu\nu}&\equiv& \frac{1}{D-3 }\left(\nabla^\rho \nabla^\sigma + \frac{D-3}{D-2}R\indices{^{\rho\sigma}}\right)W_{\rho\mu\sigma\nu}\,\label{eq:BachAp2} \\
&=&S^{\alpha\beta}W_{\mu\alpha\nu\beta} + \nabla^2S_{\mu\nu} -\nabla^\rho\nabla_{(\mu}S_{\nu)\rho}  .\label{eq:BachAp1}
\eea

The structure of nonlinear conformal gravity is quite interesting: any maximally symmetric spacetime is a solution, and the propagating degrees of freedom around these solutions form an irreducible SO$(2,4)$ representation. Around flat space, this representation is a (non-unitary) Poincar\'e representation that cannot be split apart.  Around de Sitter or Anti-de Sitter space, the representation can be split into a pair of (A)dS representations, which are that of a massless spin-2, and a partially massless (PM) spin-2~\cite{Maldacena:2011mk,Deser:2012qg}. On AdS, the PM representation is by itself non-unitary, but on de Sitter space, the reason that conformal gravity is not unitary is that there is a relative minus sign between the kinetic terms of the massless spin-2 and PM spin-2 degrees of freedom.  It is possible to project out the PM part of the representation, and obtain the wavefunction in Einstein gravity from that of conformal gravity~\cite{Maldacena:2011mk,Anastasiou:2016jix}.

\subsection{Linearized conformal gravity }

We now discuss some of the features of linearized conformal gravity. This theory can be straightforwardly obtained from the nonlinear Weyl-squared theory \eqref{eq:SconfAp} by linearizing around flat space,
\be 
g_{\mu\nu} = \eta_{\mu\nu} + 2h_{\mu\nu}\, .
\ee
The linearized Riemann and Schouten tensors are given explicitly by
\begin{align}
R_{\mu\nu\rho\sigma}^{(L)} &=-\partial_\mu\partial_\rho h_{\nu\sigma}-\partial_\nu\partial_\sigma h_{\mu\rho}+\partial_\mu\partial_\sigma h_{\nu\rho}+\partial_\nu\partial_\rho h_{\mu\sigma}\,,\\
S_{\mu\nu}^{(L)} &= \frac{1}{D-2}\left(-\partial_\mu\partial_\nu h - \square h_{\mu\nu} +2\partial^\alpha\partial_{(\mu}h_{\nu)\alpha}\right)-\eta_{\mu\nu}\frac{1}{D-1}\left(
2\partial_\alpha\partial_\beta h^{\alpha\beta} - \square h\right)\,,
\end{align}
and the linearized Weyl tensor is:
\be 
W^{(L)}_{\mu\nu\rho\sigma} = R_{\mu\nu\rho\sigma}^{(L)}-2(\eta_{\mu[\rho}S_{\sigma]\nu}^{(L)}-\eta_{\nu[\rho}S^{(L)}_{\sigma]\mu})\, .
\ee

In contrast to the nonlinear theory, the action
\be 
S =\frac{1}{4}\int \rd^Dx\, W^{(L)}_{\mu\nu\rho\sigma}W_{(L)}^{\mu\nu\rho\sigma}\, ,
\label{eq:linearconfgrav}
\ee
is invariant under both linearized diffeomorphisms and linearized Weyl transformations in {\it all} dimensions:
\begin{align}
\delta h_{\mu\nu} &= \partial_{\mu}\xi_\nu+\partial_\nu \xi_\mu\,,\\
\delta h_{\mu\nu} &= \Omega(x) \eta_{\mu\nu}\,,
\end{align}
and so it makes sense to call this linearized conformal gravity in any dimension. In fact, the linearized Weyl tensor itself is gauge invariant under both of these transformations. Much like in Einstein gravity, an important difference between the linear and nonlinear theories is that the linear theory has gauge-invariant local operators.  The linearized Weyl tensor is the basic such operator (see~\cite{Basile:2017kaz,Basile:2018eac,Joung:2021bhf} for formalisms that reveal the operator spectrum of linear conformal gravity).

The equation of motion following from~\eqref{eq:linearconfgrav} is 
\be 
\partial_\mu \partial_\rho W_{(L)}^{\mu\nu\rho\sigma}=0\, .
\label{eq:confgravEOM}
\ee
Because of the relation~\eqref{eq:WeylDiv}, which at the linearized level reads
\be 
\partial^\mu W_{\mu\nu\rho\sigma}^{(L)} = -\frac{(D-3)}{(D-2)}C_{\nu\rho\sigma}^{(L)}\, ,\label{eq:WeylDiv}
\ee
where the linearized Cotton tensor is
\be C_{\nu\rho\sigma}^{(L)}= 2(D-2) \partial_{[\sigma}S_{\rho]\nu}^{(L)}\,, \ee
we can also write \eqref{eq:confgravEOM} as the vanishing of the divergence of the linearized Cotton tensor
\be 
\partial^\rho C_{\nu\rho\sigma}^{(L)}=0\, .
\ee

Much like Einstein gravity, in $D=4$ linearized conformal gravity is electric-magnetic duality invariant, as we discussed in Section~\ref{sec:confgrav}.

It is possible to include irrelevant interaction terms in this effective field theory, paralleling the EFT of linearized Einstein gravity. Any interaction terms built out of the linearized Weyl tensor will preserve all the symmetries of the theory.
As we discuss in Section~\ref{sec:confgrav}, we can think of this EFT as being the gapless phase mandated by the anomaly~\eqref{eq:cganom}, or equivalent we can think of it as the Goldstone theory for the breaking of the electric 1-form shift symmetry of the conformal graviton.

\newpage
\section{Spectral decompositions\label{ap:Spec}}

Spectral decompositions of the current-current two-point functions play an important role in the main text. In particular, by analyzing the properties of the two-point functions that are fixed by the symmetry structure, we infer properties of the spectrum of the theory.
In this appendix we provide some additional details about the \KL decomposition, particularly in theories where the two-point functions indicates that they are not unitary. The overall strategy mirrors that of~\cite{Delacretaz:2019brr,Hinterbichler:2022agn}, where the decomposition of the current-current two-point function is used to discern the spectrum of the theory.

In this appendix we focus in particular on the case of the conformal scalar and illustrate how the spectral decomposition can be used to infer that the theory is not unitary. The general lessons can be abstracted straightforwardly to cases with spin.

\subsection{Projectors}
 This first step involves rewriting the two-point function \eqref{eq:TwoPFConfScalar} in terms of projectors that split up the various components of the two-point function. These projectors are
 defined in terms of the simpler objects: 
\be 
\pi^{(0)}_{\mu\nu} = \frac{p_\mu p_\nu}{p^2}\, ,\qquad \qquad \pi^{(1)}_{\mu\nu} = \eta_{\mu\nu}-\frac{p_\mu p_\nu}{p^2}\, .
\ee
A basis of projectors is given by
\begin{align}
\Pi^{(0)}_{\mu_1\mu_2\nu_1\nu_2} &= \frac{D}{D-1}\left(\pi^{(0)}_{\mu_1\mu_2}-\frac{\eta_{\mu_1\mu_2}}{D}\right)\left(\pi^{(0)}_{\nu_1\nu_2}-\frac{\eta_{\nu_1\nu_2}}{D}\right)\, ,\label{eq:proj1}\\
\Pi^{(1)}_{\mu_1\mu_2\nu_1\nu_2} &= \frac{1}{2}\left(\pi^{(0)}_{\mu_1\nu_1}\pi^{(1)}_{\mu_2\nu_2}+\pi^{(0)}_{\mu_1\nu_2}\pi^{(1)}_{\mu_2\nu_1}+\pi^{(1)}_{\mu_1\nu_1}\pi^{(0)}_{\mu_2\nu_2}+\pi^{(1)}_{\mu_1\nu_2}\pi^{(0)}_{\mu_2\nu_1}\right)\, ,\\
\Pi^{(2)}_{\mu_1\mu_2\nu_1\nu_2} &= \frac{1}{2}\left(\pi^{(1)}_{\mu_1\nu_1}\pi^{(1)}_{\mu_2\nu_2}+\pi^{(1)}_{\mu_1\nu_2}\pi^{(1)}_{\mu_2\nu_1}-\frac{1}{D-1}\pi^{(1)}_{\mu_1\mu_2}\pi^{(1)}_{\nu_1\nu_2}\right)\label{eq:proj2}\, ,
\end{align}
which satisfy the orthonormality conditions
\be 
\Pi^{(i)}_{\mu_1\mu_2\rho_1\rho_2}\Pi\indices{^{(j)\rho_1\rho_2}_{\nu_1\nu_2}} = \delta^{ij}\Pi^{(i)}_{\mu_1\mu_2\nu_1\nu_2}\, ,
\ee
with $i,j=0,1,2$, and the completeness relation 
\be 
\Pi^{(0)}_{\mu_1\mu_2\rho_1\rho_2}+\Pi^{(1)}_{\mu_1\mu_2\rho_1\rho_2}+\Pi^{(2)}_{\mu_1\mu_2\rho_1\rho_2} = \frac{1}{2}\left(\eta_{\nu_1\mu_2}\eta_{\mu_1\nu_2} + \eta_{\mu_1\nu_1}\eta_{\mu_2\nu_2}\right)-\frac{1}{D}\eta_{\nu_1\nu_2}\eta_{\mu_1 \mu_2}\, ,
\ee
where the tensor on the right-hand side
is the identity on the space of symmetric traceless tensors. We can think of $\Pi^{(i)}$ as the projector onto the spin-$i$ representation within this space.

\subsection{Spectral decomposition}

We now argue that the conformal scalar theory is non-unitary using the spectral decomposition of the two-point function~\eqref{eq:TwoPFConfScalar}. This correlator is easily rewritten in terms of the projectors~\eqref{eq:proj1}--\eqref{eq:proj2}. This yields 
\be 
\braket{J_{\mu\nu}K_{\rho\sigma}} = -\left(\Pi^{(1)}_{\mu\nu\rho\sigma}+ \Pi^{(2)}_{\mu\nu\rho\sigma}\right)\, .\label{eq:ToMatchAp1}
\ee

Our starting point is the spectral decomposition for a correlator of two symmetric traceless tensors 
\be
\braket{{J}_{\mu\nu}K_{\rho\sigma}} 
= \int_0^\infty\, \rd s\, \frac{s}{p^2 +s} \left(\rho_0(s)\tilde{\Pi}^{(0)}_{\mu\nu\rho\sigma}+ \rho_1(s)\tilde{\Pi}^{(1)}_{\mu\nu\rho\sigma}+ \rho_2(s)\tilde{\Pi}^{(2)}_{\mu\nu\rho\sigma}\right)\, .
\label{eq:galKLdecomp}
\ee
Here $\rho_j(s)$ are the spin $j$ components of the spectral density (the only massless representation that can couple to a symmetric conserved current is a scalar, so the spectral densities of the spin-1 and spin-2 states must go to zero as $p^2\to 0$). Implicit in the spectral decompostion~\eqref{eq:galKLdecomp} is the assumption that the representations contributing to the correlator have propagators that go like $p^{-2}$~\cite{Weinberg:1995mt}, which is satisfied by all unitary Poincar\'e representations. As usual~\cite{Hinterbichler:2022agn}, the tensors $\tilde{\Pi}^{(i)}$ that appear in \eqref{eq:galKLdecomp} are not quite the projectors ~\eqref{eq:proj1}--\eqref{eq:proj2}, but are instead off-shell versions of the projectors that are obtained by replacing $p^2\rightarrow -s$ in ~\eqref{eq:proj1}--\eqref{eq:proj2}. Hence, they depend on $s$ and reduce to the projectors when $s \to -p^2$.

In order to perform the spectral decomposition, the goal is to match~\eqref{eq:galKLdecomp} to~\eqref{eq:ToMatchAp1}, and solve for the three spectral densities $\rho_j$. In this case, there is no choice of the $\rho_j(s)$ that works. At its core, this is because we are trying to match a two-point function that goes like $p^{-4}$ with a sum of propagators that go like $p^{-2}$. Since all unitary representations have propagators that scale like this, this is already enough to see that the theory is not unitary. Of course, we could generalize the \KL representation to allow for totally generic behavior of the Feynman propagator of the exchanged representations (and therefore allow all possible non-unitary irreducible representations to be exchanged as well). In this case it would be possible to match the two point function and extract the spectrum of the theory. It would actually be quite interesting to do this, as the properties of non-unitary Poincar\'e representations are not particularly well understood. But for our purposes, it suffices to conclude that there must be non-unitary representations in the theory.

Not that
the structure of anomalies completely fixes the nonlocal part of the current-current two-point function, so the structure of these poles is intrinsic to the anomaly. There is an ambiguity about where to put the anomaly in a conservation equation, but this freedom only changes the contact terms appearing in the correlator, i.e. the terms analytic in $p^\mu$, and so cannot modify the structure of the poles.

Note that the essential tension has nothing to do with the spin of the fields involved, it was merely the difficulty of matching a two-point function that scales like $p^{-4}$ to a spectral decomposition where each of the constituents have a propagator that scales like $p^{-2}$. These same ingredients are present in the conformal gravity case. There, the two-point function scales like $p^{-4}$, and it is similarly impossible to match it to the \KL decomposition under the assumption that all of the states have ordinary $p^{-2}$ propagators (which all unitary representations do). We can therefore conclude that the theory is non-unitary also in that case.

\newpage
\linespread{.975}
\addcontentsline{toc}{section}{References}
\bibliographystyle{utphys}
{\small
\bibliography{ConformalGravity-arxiv}
}

\end{document}